\documentclass[12pt,preprint]{aastex}

\usepackage{natbib,graphics,epsfig}

\slugcomment{Accepted for publication in ApJ}

\shorttitle{FUSE Observation of the Narrow-line Seyfert 1 Galaxy RE 1034$+$39}
\shortauthors{Casebeer, Leighly and Baron}

 \newcommand{\ew}{$W_{\lambda}$ }

\newcommand{\ovi}{\ion{O}{6} $\lambda1034$ }
\newcommand{\heii}{\ion{He}{2} $\lambda1640$ }
\newcommand{\oiv}{\ion{O}{4} $\lambda1402$ }

\newcommand{\re}{RE~1034$+$39 }
\begin{document}

\title{FUSE Observation of the Narrow Line Seyfert 1 Galaxy RE~1034$+$39:
Dependence of Broad Emission Line Strengths on the Shape of the Photoionizing
Spectrum\footnote{Based on observations made with the NASA-CNES-CSA
  Far Ultraviolet  Spectroscopic Explorer. FUSE is operated for NASA
  by the Johns Hopkins  University under NASA contract NAS5-32985.}} 

\author{Darrin A.\ Casebeer, Karen M.\ Leighly,  and E.\ Baron}
\affil{Department of Physics and Astronomy, The University of
  Oklahoma, 440 W. Brooks St., Norman, OK 73019}
\email{casebeer@nhn.ou.edu}

\begin{abstract}
We present analysis from simultaneous {\it FUSE}, {\it ASCA}, and {\it
EUVE} observations, as well as a reanalysis of archival {\it HST}
spectra, from the extreme Narrow-line Seyfert 1 Galaxy RE~1034$+$39
(KUG~1031$+$398).  RE~1034$+$39 has an unusually hard spectral energy
distribution (SED) that peaks in the soft X-rays.  Its emission lines
are unusual in that they can all be modelled as a Lorentzian centered
at the rest wavelength with only a small range in velocity widths.  In
order to investigate whether the unusual SED influences the emission
line ratios and equivalent widths, we present three complementary
types of photoionization analysis.  The {\it FUSE} spectrum was
particularly important because it includes the high-ionization line
\ion{O}{6}. First, we use the photoionization code {\it Cloudy} and
the SED developed from the coordinated observations to confirm that
the emission lines are consistent with observed hard SED. The best
model parameters were an ionization parameter $\log(U) \approx -2$ and
a hydrogen number density $\log(n_H)=9.75$ [$\rm cm^{-2}$].  Second, we present a
Locally Optimally-emitting Cloud model.  This model produced enhanced
\ion{O}{6} as observed, but also yielded far too strong \ion{Mg}{2}.
Third, we develop a series of semi-empirical SEDs, run {\it Cloudy}
models, and compare the results with the measured values using a
figure of merit (FOM).  The FOM minimum indicates similar SED and gas
properties as were inferred from the one-zone model using the
RE~1034$+$39 continuum. Furthermore, the FOM increases sharply toward
softer continua, indicating that a hard SED is required by the data in
the context of a one-zone model.
\end{abstract}

\keywords{line: formation --- quasars: emission lines --- quasars:
  individual (KUG 1031$+$398)}

\section{Introduction}

The broad emission lines in Active Galactic Nuclei (AGN) are powered
primarily by photoionization (e.g., Osterbrock 1989). There is copious
evidence leading to this conclusion.  For example, the emission lines
are observed to vary in response to variations in the continuum.
Also, the flux in strong lines is correlated with the continuum flux.
Knowing that photoionization powers the emission lines, we can easily
predict which lines will be strong.  Photoionization removes electrons
from hydrogen atoms; therefore, recombination products such as
Ly$\alpha$ should be strong. The free electrons collide with ions from
other elements, exciting those from abundant elements that are
isoelectronic with lithium, such as C$^{3+}$, because the excitation
potential of their first excited states is low.  Many of these excited
states are short-lived; the ion returns to ground state with emission
of a photon which then escapes the gas, cooling it. Thus, we expect
lines such as \ion{C}{4}~$\lambda 1549$ to be strong as well.  

For this cooling channel to be effective, three-times ionized carbon
atoms must be present in the gas.  Therefore, the line emission should
be governed by the photoionization parameter $U=\Phi/nc$ where $\Phi$
is the photoionizing flux, generally in units of $\rm photons\,
cm^{-2}\,s^{-1}$, and $n$ is the gas density in units of $cm^{-2}$.  The ionization
parameter is the ratio between the photon density, which drives
photoionization, and the gas density, which controls recombination.
Thus, since \ion{C}{4} being strong implies the presence of C$^{3+}$ in the
gas, it also implies a particular value or range of values for the
ionization parameter (more specifically, the ratios of Ly$\alpha$ to
\ion{C}{4} and of \ion{C}{4} to \ion{C}{3}).  A typical
value of the ionization parameter inferred is around $\log(U)=-1.5$
(e.g., Kwan \& Krolik 1981).

But does the presence of \ion{C}{4} really imply a particular
ionization parameter?  Perhaps it really is telling us that gas with a
range of ionization parameters, corresponding to a range of photon
fluxes and densities, is present, but gas with the ionization
parameter suitable for \ion{C}{4} production happens to emit most
efficiently, due to the easy excitation of the C$^{3+}$ ion.  This would
cause the integrated line emission to be weighted toward gas with
copious C$^{3+}$ present, but would not preclude the presence of less
effectively-emitting gas.  These ideas were proposed by Baldwin et
al.\ (1995) as the Locally Optimally-emitting Cloud model. 

This simple conceptual picture of photoionization in AGN provides a
zeroth-order understanding of why the line emission is similar in all
objects, and why the same strong lines are usually seen.  However,
there are some significant trends in the emission line properties
that cannot be explained directly using this zeroth-order
understanding.  Furthermore, there are numerous
examples of objects or classes of objects whose emission lines differ
from that of the average quasar.  These trends and the presence of
classes of unusual objects require more sophisticated models than the
zeroth-order picture presented above.   This is a good thing; there
are many fundamental things we don't know about the broad-line region
even after years of study, including the geometry, the dynamics of the
gas, and the origin of the gas.  With such a large number of potentially
variable parameters, our task may seem hopeless.  That is far from the
case, and in fact, the potential for solving this problem is orders of
magnitude higher than it was when the first models were being
developed due to the availability of powerful computers,
increasingly-sophisticated models, and large samples of high-quality
spectra from, e.g., the Sloan Digital Sky Survey.  

One of the factors that may affect the emission lines in AGN is the
shape of the spectral energy distribution (SED).  
AGN are known to have broad spectral energy distributions, with almost
equal power emitted per logarithmic frequency range between the
infrared and X-ray bands (e.g., Peterson 1997).  Their emission lines
are powered largely by the bandpass from $\sim 4$ eV through the soft
X-rays, although the regions of the spectrum outside this range can
contribute through heating.  This part of the spectral energy
distribution is thought to be primarily the emission of the accretion
disk, with the higher energies above $\sim 1\rm\, keV$ or so emitted
by a Comptonizing corona.  Theoretically, the continuum emitted by the
accretion disk should depend on the mass and accretion rate (e.g.,
Frank, King, \& Raine 2002; Ross, Fabian, \& Mineshige 1993), the
accretion disk geometry (e.g., Zdziarski \& Gierlinski 2004), and
inclination (e.g., Laor \& Netzer 1989; Puchnarewicz et al.\ 2001).
In the simplest thin disk models, accretion onto smaller black holes,
and accretion at a higher rate relative to the Eddington value
predicts hotter disks (e.g., Shakura \& Sunyaev 1973), and
correspondingly overall harder spectral energy distributions.  The
X-ray emission arising from Comptonization by a hot optically-thin
component may also be affected by the accretion rate as the power for
the hot component must ultimately come from accretion.  In some
models, all of the accretion energy is deposited initially into the
corona and then a fraction is reprocessed by the disk (Svensson \&
Zdziarski 1994). Other models propose an accretion/evaporation
scenario in which the disk and corona are coupled (R\'ozanska \&
Czerny 2000a,b).  In these models, the amount of emission from the
corona relative to that from the disk depends on the accretion rate
relative to the Eddington value, with the coronal emission becoming
less important at high accretion rates (e.g., Bechtold et al.\ 2003).
In addition, at very low accretion rates, the accretion should have a
different geometry (e.g., Ho 2003); a hot, optically-thin thermal
plasma may dominate the inner region, producing a much different
spectral energy distribution that lacks a big blue bump.

Thus, there is a clear theoretical expectation that the spectral
energy distributions in AGN should take on a range of shapes.  What is
the observational evidence that such a range exists?  First, it is
worth noting that there are difficulties in measuring the intrinsic
spectral energy distribution.  Since AGN vary at all wavelengths,
observations in different wavebands should in principle be
at least contemporaneous.  Absorption and reddening in the host galaxy
or in the AGN itself poses another serious complication.  Finally,
because of absorption, the portion of the observed-frame spectrum
between $13.6$ and $\sim 100$ eV cannot be observed at all.  

Nevertheless, there is some solid evidence AGN are not characterized
by a single spectral energy distribution; a range in shapes
exists. For example, studies of samples of objects shows a range of
SEDs (e.g., Elvis et al.\ 1994).  In addition, $\alpha_{ox}$, the
point-to-point slope between 2500\AA\/ and 2~keV, has been repeatedly
observed to depend on the UV luminosity (Wilkes et al.\ 1994; Bechtold
et al.\ 2003; Vignali et al.\ 2003).    Among objects in a sample of
soft X-ray selected AGNs, a correlation between the optical/UV
continuum slope and the X-ray photon index has been found, such that
objects with bluer optical continua have steeper X-ray spectra (Grupe
et al.\ 1998).  Also, a {\it FUSE} composite spectrum is observed to
be harder than an {\it HST} composite over the same rest bandpass; as
the {\it FUSE} spectra are from relatively lower luminosity objects,
this result is attributed to luminosity-dependent evolution of the
spectral energy distribution (Scott et al.\ 2004).

Indirect evidence for a range of spectral energy distributions among
AGNs is provided by the observation that the ratio of optical depth of
\ion{He}{2} Ly$\alpha$ to that \ion{H}{1} Ly$\alpha$ in the era of
\ion{He}{2} reionization fluctuates on small redshift scales.  This is
inferred to be caused by the response of these ions to a range in
continuum shapes due to the difference in ionization potentials
(Shull et al.\ 2004.)  However, at least part of this range in
continua may certainly be caused by reddening/absorption.

Emission lines in AGNs, while broadly similar, clearly show trends and
dependencies that are plausibly dependent on the spectral energy
distribution.  The first trend discovered was the Baldwin effect,
which is the empirical inverse relationship of emission line (e.g.,
\ion{C}{4}) equivalent width with luminosity (Baldwin 1977; for a
review, Osmer \& Shields 1999).  Initially, several models were
proposed for this effect.  For example, Mushotzky \& Ferland (1984)
posited a luminosity-dependent ionization parameter; a
luminosity-dependent covering fraction was also thought to be possibly
important.  More recently, observational evidence has been repeatedly
found that the slope of the Baldwin effect is steeper for
high-ionization lines than for low-ionization lines (Osmer, Porter, \&
Green 1994; Green 1996; Espey \& Andreadis 1999; Wilkes et al.\ 1999;
Green, Forster, \& Kuraszkiewicz 2001; Dietrich et al.\ 2002; Shang et
al.\ 2003; Warner et al.\ 2004).  This is interpreted appealingly as
evidence that, as the luminosity increases, the spectral energy
distribution softens, resulting in a more rapid decrease in the highest
ionization lines.  Further evidence supporting this view comes from
observed correlations between continuum shape indicators such as
$\alpha_{ox}$ and line equivalent width (Zheng, Kriss \& Davidsen
1995; Wang et al.\ 1998; Green 1998; Wilkes et al.\ 1999; Scott et
al.\ 2004).  Thus, it appears that the Baldwin effect is a phenomenon
that depends principally on the shape of the spectral energy
distribution, and its corresponding dependence on luminosity.

Interestingly, it has been found that Narrow-line Seyfert 1 galaxies
show a Baldwin effect that is offset from that shown by other Seyferts
and quasars (Wilkes et al.\ 1999; Leighly \& Moore 2004;
Leighly et al.\ in prep.).  This might be related to the fact that the cores
of the lines show a stronger Baldwin effect than the wings of the
lines (Osmer, Porter \& Green 1994; Shang et al.\ 2003; Warner et al.\
2004).  It is possibly an indication of a different BLR geometry for
this class of objects (e.g., Leighly 2004).  

Theoretical confirmation of this model for the Baldwin effect is
promising but not yet completely firm.  Korista, Baldwin \& Ferland
(1998) use an LOC model and find that the slope dependence of the data
is consistent with an evolving spectral energy distribution.  On the
other hand, Korista, Ferland \& Baldwin (1997) show that there may be
difficulties explaining the strength of the \ion{He}{2} line using
continua developed from composite spectra.  

The second important trend for line emission in AGN is the Eigenvector
1 dependence.  Boroson \& Green (1992), in a principle
components study of spectra from PG quasars, discovered that much of
the variance among optical emission lines is correlated such that
narrow lines are associated with strong \ion{Fe}{2} and weak
[\ion{O}{3}] emission.  Later work shows a relationship with the
steepness of the X-ray spectrum (e.g., Brandt \& Boller 1999), and an
extension to the UV spectra (e.g., Wills et al.\ 1999).  The physical
driver for Eigenvector 1 is thought to be the accretion rate relative
to Eddington (e.g., Leighly \& Moore 2004 and references in the
introduction). Precisely how this works has not been yet clearly
delineated, but it could be mediated by the spectral energy
distribution.  For example, Wandel \& Boller (1998) propose that the
steeper soft X-ray spectrum seen in NLS1s translates to a more
powerful extreme UV excess that would cause the emission lines to
originate from a larger radius where velocities are
smaller. Kuraszkiewicz et al.\ (1999) attribute the characteristic UV
emission-line ratios to higher densities in the broad-line region.
They examine a two-phase BLR model, and suggest that the higher
densities arise naturally in this model due to the hotter big blue
bumps. Wills et al. (1999) infer high densities in the BLR as well,
from the \ion{Si}{3}]/\ion{C}{3}] ratio.  However, as shown in \S 7.3,
this ratio can depend on the spectral energy distribution, because in
some regions of the U--SED-shape parameter space, \ion{Si}{3}] will
  dominate the cooling. 

The spectral energy distribution can influence the line properties
in another way.  Leighly \& Moore (2004) and Leighly (2004) show that
in a sample of Narrow-line Seyfert 1 galaxies, the degree of blueshift
of  \ion{C}{4} is correlated with $\alpha_{ox}$, such that
objects that are relatively weak in X-rays have more highly
blueshifted lines.  This makes perfect sense in the context of
radiative-line driven wind models (e.g., Proga, Stone \& Kallman
2000), because the way that the UV and X-rays play competing roles in
the formation of such winds.  Specifically, the UV accelerates the
wind, while the X-rays tend to impede the wind because they can
overionize it, destroying the ions that scatter the UV
emission.  So, even though objects with relatively strong X-rays can
have winds, due to self-shielding, or preabsorption of soft X-rays
before reaching the wind gas (Murray et al.\ 1995; Proga \& Kallman
2004), objects with relatively strong UV emission and weak X-ray
emission will always drive faster, more massive winds.  

Leighly (2004) also notes that the spectral energy distribution can
influence the strengths of the emission lines indirectly via the wind.
It has long been suspected that low- and intermediate-ionization lines
are produced in a different region than high-ionization lines.  The
evidence for this comes from modeling of the line emission (e.g.,
Collin-Souffrin et al.\ 1982, 1986; Wills, Netzer, \& Wills 1985), and
from the difference in profiles (e.g., Collin-Souffrin et al.\ 1988;
Baldwin 1997).  One of the puzzling features in quasar spectra is the
observed trend for objects with strong high-ionization line emission
such as \ion{N}{5} to have strong low-ionization line emission such as 
\ion{Si}{2} (Wills et al.\ 1999).  Leighly (2004) show that
low-ionization line emission can be enhanced if the continuum
illuminating that emission region is first filtered through the
wind. Such a continuum is deficient in helium-continuum photons, as
they are all absorbed by the wind, and thus can only excite relatively
low-ionization species.

We are currently investigating the effects of the spectral energy
distribution on the broad-line emission and kinematics in AGN.  In
this installment, we present an analysis of spectra from a Narrow-line
Seyfert 1 galaxy RE~1034$+$39 ($z=0.043$). This object is well known
for its unusual spectral energy distribution.  Puchnarewicz et al.\
(1995), Puchnarewicz, Mason, \& Siemiginowska (1998), and Puchnarewicz
et al.\ (2001) have reported optical, UV and X-ray observations of
this object.  While the optical and UV spectra of the typical AGN
rises toward the blue, forming the big blue bump, RE~1034$+$39's
optical/UV spectrum is rather red.  The big blue bump appears instead
in the X-ray spectrum: it appears to peak at $\sim 100 \rm \, keV$,
and the high energy turnover toward long wavelengths appears to be in
the soft X-rays (Puchnarewicz et al.\ 1995).  This unusual spectral
energy distribution was confirmed by the {\it BeppoSAX} observation
conducted in 1997 and reported by Puchnarewicz et al.\ (2001).  The
unusual spectral energy distribution has been modelled as an edge-on
accretion disk (Puchnarewicz et al.\ 2001) and as an irradiated
accretion disk (Soria \& Puchnarewicz 2002).

Given RE~1034$+$39's hard spectral energy distribution, one might
expect that there would be strong high-ionization line emission in its
UV spectrum, assuming that the line emitting gas sees the same
continuum that we do. To test this hypothesis, we obtained a {\it FUSE}
observation of RE~1034$+$39, along with coordinated {\it EUVE} and
{\it ASCA} X-ray observations.  We present the analysis of these
spectra here, as well as a reanalysis of the archival {\it HST}
spectrum (\S 2).  In the next three sections, we present three
complementary types of photoionization analysis.  
First, using the spectral energy distribution constructed
from the coordinated observations, we ran {\it Cloudy} (Ferland 2001)
photoionization models to determine the conditions of the line
emitting region, and ascertain whether the line equivalent widths and
ratios are consistent with the extreme spectral energy distribution
for a single-zone model (\S 3). We then look at how well the
lines could be modeled in the LOC scenario (\S 4).  Third, we consider a
range of semi-empirical spectral energy distributions presented in the
Appendix, and compare the predicted line emission as a function of SED
(and other parameters) with that of RE~1034$+$39 (\S 5).  We
discuss the results in \S 6, and present auxiliary material in an
Appendix, \S 7.

\section{Multiwavelength Observations and Data Analysis}

\subsection{{\it FUSE} Observation}

The {\it Far Ultraviolet Spectroscopic Explorer
(FUSE)}\footnote{http://fuse.pha.jhu.edu/} was launched 1999 June 24.
{\it FUSE} provides unprecedented sensitivity and resolution over the
905--1187\AA\/ bandpass, which contains rest wavelength
transitions from many important atoms and molecules (Moos et al.\
2000; Sahnow et al.\ 2000).  The {\it FUSE} observation of
RE~1034$+$39, coordinated with observations using {\it ASCA} and {\it
EUVE}, was approved as part of the Cycle 1 Guest Investigator
observing program. 

The observing log for the {\it FUSE} observation is given in Table~1.
The raw {\it FUSE} data were downloaded and reprocessed.  At the time
that the analysis was done, CALFUSE v2.0 had not yet been released,
and the Beta version was not fully functional; therefore, we used
CALFUSE v1.8.7 with some enhancements from v2.0 as described below.
We recently re-ran the pipeline using CALFUSE v2.4 and confirmed
that the resulting spectra are indistinguishable from the ones that
are presented here.  RE~1034+39 is a faint object for {\it FUSE};
therefore, we used only the data obtained during the time that {\it
FUSE} spent in the earth's shadow.  To reduce the background, we only
used data within PHA channels 4--16 (e.g., Brotherton et al.\ 2002).
This choice approximately halved the value of the constant background
component, but decreased the source flux by only $\sim 5$\%.

CALFUSE v1.8.7 approximated the background using a uniform image;
however, the real detector background has considerable structure.  In
CALFUSE v2.0 and subsequent versions, this is accounted for by using
real images of the background.  In order to utilize this improvement
in CALFUSE v1.8.7, we extracted the night-time background image from
the v2.0 calibration files and used it instead of the v1.8.7 background
image.  We then ran the pipeline again.  We adjusted the
scaling and offset of the background image to the background level of
the observation.  To do this, we extracted spectra from two
source-free regions of the detector in the observed image and the
background image.  A linear regression between the background and
source vectors produced the scale and offset necessary to match the
background file to the data.  The background was suitably scaled and
offset, and the pipeline script modified to use this background image.

We examined the two-dimensional images after background subtraction.
The spectral trace was clearly visible for all telescope systems
except SiC2.  We take this as evidence that the image of the galaxy
was not in the the slit for this telescope system, and we do not consider it
further.  The background subtraction appeared to be successful for
most of the telescope/detector segments; however, we were unable to
obtain a satisfactory background subtraction for the LiF2b spectrum,
and we do not consider it further.  The SiC1a spectrum was not used in
the composite spectrum, due to low signal-to-noise ratio compared to
the LiF1a spectrum which samples the same wavelength interval.

We examined the flux calibration between overlapping detector
segments. The most noticeable difference was between LiF2a and LiF1b
segments longward of 1140 \AA\/.  We expect that this difference may
be due to the ``worm'' or shadowing of the detector by the grid wires
above it in LiF1b.  Therefore, we exclude data longward of 1140 \AA\/
for that segment, and average the two segments over the 
remainder of the overlapping interval.

Wavelength calibration V10 was used.  The remaining spectra (LiF1a,
combined LiF1b and LiF2a, SiC1b) were rebinned using an algorithm
similar to the IRAF {\tt trebin} script.  The spectra were then
concatenated, and the result is shown in Fig.\ 1, after having
corrected for the redshift ($\rm z=0.043$), and Galactic reddening
$E(B-V)=0.015$ mag, estimated from the infrared cirrus (Schlegel,
Finkbeiner \& Davis 1998), using the Cardelli, Clayton \& Mathis
(1989) reddening curve.  

Uncertainties in spectra are rarely shown for {\it FUSE} spectra of
AGN (e.g., Shang et al.\ 2005; Romano et al.\ 2002).  This is possibly
because the error column in the pipeline spectra aren't very useful;
the spectra contain only several photons per bin at full wavelength
resolution.  However, uncertainties on the spectra are useful to have
for faint objects such as AGN, so we developed a method to evaluate
them.  First, we ran the pipeline, outputting the intermediate
products which include the ``total'' image ({\it \_img.fit}) and the
scaled background image ({\it \_bkgd.fit}).  The spectra at full
wavelength resolution were extracted from these images using the
nominal extraction regions (calibration files {\it spex1a009.fit},
etc) and collapsing in the y direction.  The wavelength scale
was obtained from the pipeline output spectrum.  The total counts
spectrum and background counts spectrum were rebinned to the desired
binsize (in this case, 0.5\AA\/) by summing the counts within each
bin.  Then the error counts spectrum was obtained by propagation of
errors assuming Poisson statistics in each bin of the rebinned total
and background spectra.  The error spectrum was then ``fluxed'' using
the appropriate calibration flux file (calibration files {\it
  flux1a008.fit}, etc.).  The error spectrum is shown by the thick
line in Fig.\ 1.  This shows that no continuum flux was detected below
$\sim 960$\AA\/ where only the SiC1b spectrum is available.

The shorter wavelength component of the \ovi doublet (1031.93\AA\/) is
clearly detected, but the longer wavelength component (1037.62\AA\/)
is absent (Fig.\ 1).  We suspect that the longer wavelength component
does not appear because of residual background subtraction problems.
Fig.\ 2 shows a section of the 2D LIF1 image after screening and
before background subtraction.  The spectral trace is clearly seen,
and an enhancement corresponding to the 1037\AA\/ component is
apparent.  Fig.\ 2 also shows that the background is higher near the
edge of the detector; the effective area is lower as well.  Thus it
proved to be impossible to tease the longer wavelength component of
\ion{O}{6} from the noise.  We note that there is an enhancement in
the low signal-to-noise ratio Sic1a spectrum at 1037\AA\/ (shown as
an inset in Fig.\ 1). Therefore,
we base our analysis on the 1032\AA\/ component. 

\ion{C}{3} $\lambda 977$ was also detected, as well as Ly$\beta$ (see
Section 2.6); however, \ion{N}{3} and \ion{He}{2}, low equivalent
width lines sometimes seen in other FUV spectra from AGN (e.g.,
Brotherton et al.\ 2002; Hamann et al.\ 1998) are were not detected. 

\subsection{ASCA Observation}

An {\it ASCA} observation was conducted simultaneously with the {\it FUSE}
observation (Table 1).  A standard configuration was used during the
observation.  The GISs were operated in PH mode throughout the
observation.  The SISs were operated in 1~CCD  Faint and Bright mode
at high and medium bit rate, respectively.  The  SIS energy gain was
reprocessed using the latest calibration file ({\sf
  sisph2pi\_290301.fits}). 

We used the same criteria as the  rev~2 standard
processing for our data
selection\footnote{http://adfwww.gsfc.nasa.gov/asca/processing\_doc/proc/rev2/la  
test/screen.html}, except for the {\sf ANG\_DIST} (angular distance
from the nominal pointing) criterion.  The pointing accuracy was poor
during this observation; therefore, we used a looser angular distance
selection criterion. The sensitivity change due to the pointing error
of 1.0~arcmin is about 10\%; this is accounted for in the spectral
fitting by the ancillary response file.

The source was detected at R.A.$={\rm 10^h 34^m 36^s}$ and
Decl.$=39^\circ 38{'} 56^{"}$ (equinox 2000.0) in good agreement with
the optical position within the {\it ASCA} position accuracy of
$\sim1^{'}$.  We accumulated the source photons from the circular
regions centered on the source with radii of $4.0$, $3.5$ and
$6.0$~arcmin for the SIS-0, SIS-1 and GISs, respectively.  The
background data were accumulated from source-free regions on the
detector.  The average count rates from the source regions are
0.096~c~s$^{-1}$ and 0.059~c~s$^{-1}$, and the background fractions are
estimated to be $\sim$13\% and $\sim$33\% for each SIS and GIS,
respectively.

We fitted each of the four spectra separately with a model
consisting of a power-law with Galactic absorption
($N_H=1.43\times10^{20}\rm \,cm^{-2}$; Murphy et al.\ 1996) in the
whole energy band. The statistics were rather poor in this short
observation; thus, this model is not rejected at 90\% confidence
level for any of the spectra.  Then, we added a blackbody to the model
to represent the soft excess component commonly seen in NLS1 spectra,
and previously seen in this object (Pounds, Done \& Osborne 1996;
Leighly 1999).  The fits were acceptable for all four spectra and
$\chi^2$ values were improved significantly (F-test confidence levels
of 99 and 97\% for SISs and GISs, respectively).  All parameters are
consistent among the four detectors.  We also confirmed that the results
from the faint and bright mode are consistent with those only from the
faint mode data (which have better energy resolution).  Finally we fit
the SISs and GISs data simultaneously with a model of a power-law plus
a black body attenuated by Galactic absorption.  The best-fit
parameters are summarized in Table~\ref{xray}.

The divergence of the SIS spectra at low energies has been reported,
especially for observations performed near the end of the mission.  An
empirical correction for this has been proposed by Yaqoob et al.\
(2000)\footnote{http://heasarc.gsfc.nasa.gov/docs/asca/calibration/nhparam.html}. 
We applied this correction by introducing a fixed neutral absorption
$N_H =7.3\times10^{20}\rm \,cm^{-2}$, appropriate for the date of
this observation, and discarding the SIS-1 data below 1~keV.  The
best-fit parameters from this fit are also listed in Table 3. 

\subsection{EUVE Observation}

An {EUVE} observation was conducted simultaneously with the {\it ASCA}
and {\it FUSE} observations (Table 1).  A description of the standard
reduction for the {\it EUVE} Deep Survey Instrument is given in
Halpern, Leighly \& Marshall (2003).  RE~1034+39 is a reasonably
bright target for {\it EUVE}. However, the observation was conducted
during Solar Maximum, which strongly enhanced the background and
resulted in noisy data.  While the resulting light curve was formally
significantly variable ($\chi^2_R$ for a constant hypothesis was 1.99
for 53 degrees of freedom), we believe that the origin of the
variability may be less-than-perfect background subtraction.  We note
that the average count rate observed ($0.036\rm\, count\,s^{-1}$)
during the coordinated observation is consistent with the rate
observed ($0.037\rm\, count\,s^{-1}$) during a quieter period three
years earlier and is therefore consistent with the general lack of
variability in this object (Puchnarewicz et al.\ 2001).

We use PIMMS (Portable, Interactive Multi-Mission Simulator;
Mukai 1993) to estimate the flux.  We assume that the spectrum is
a power law with Galactic absorption of $N_H=1.43 \times 10^{20}\rm \,
cm^{-2}$.  For a range of photon indices between 2.5--4.5, we predict
an intrinsic flux at 0.15\,keV observed frame of between
0.75--0.83$\,\rm photons\,cm^{-2}\,s^{-1}\,keV^{-1}$.

\subsection{Archival HST FOS spectra}

Our proposal for coordinated {\it HST} observations was not
successful, so we analyzed the archival data from this object,
previously presented by Puchnarewicz et al.\ (1998).  We applied the most
up-to-date  calibration to the archival data.  We used the Galactic
absorption lines to check the wavelength calibration of the spectra
and shifted them appropriately. 

The {\it HST} data were obtained more than three years before the {\it
FUSE} observations.  Thus, variability between the observations
was a concern; are we justified in using these archival data?  We
believe that we are,  for two reasons. First, the {\it
FUSE} and {\it HST} continua are separated by a gap of less than 20
\AA, and the {\it FUSE} continuum appears to be consistent with an
extrapolation of the {\it HST} continuum.  Second, \re is well known
for its general lack of variability (Puchnarewicz et al.\ 2001).

\subsection{Analysis of the Emission Lines}

In this section, we describe our extraction of the emission-line
parameters from the spectra, and discuss some features of the
results.  The spectra were fitted using the {\tt Specfit} module in
IRAF (Kriss 1994).  The results are
given in  Table 3 and  Fig.\ 3.

Puchnarewicz et al.\ (1998) found no evidence for intrinsic absorption
lines in the {\it HST} spectra; our reanalysis did not uncover any
either.  The high resolution and good sensitivity obtained in the
{\it FUSE} spectrum revealed a number of weak absorption features on
\ion{O}{6} (Fig.\ 4). The wavelengths of all of these features were
consistent with Galactic molecular hydrogen absorption.  Thus, we find
no evidence for intrinsic \ion{O}{6} absorption in RE~1034$+$39. 

Preliminary examination showed that the spectra consisted of a smooth
continuum and narrow lines that are generally not strongly blended.
Also, RE~1034$+$39 lacks the strong \ion{Fe}{2} multiplets frequently found
in Narrow Line Seyfert 1 galaxies that can interfere with spectral
fitting.   Therefore, we modeled the continuum locally using a linear
function  around each line or blend.  The line profile of the
permitted and semi-forbidden lines could be modeled well with  a
Lorentzian profile and there appeared to be little evidence for a
difference in shape between the high and low-ionization lines such as
has been observed in some Narrow Line Seyfert 1 galaxies (e.g.,
Leighly \& Moore 2004a).  In fact, we found that we could obtain an 
acceptable fit if the wavelength of each line were fixed to the
laboratory wavelength, allowing only the width and flux to vary.
Widths of doublets and multiplets were constrained to be equal.
Doublet and multiplet intensities of some lines were constrained 
as follows.  The relative line intensities for \oiv multiplet were
constrained using the treatment described by Nussbaumer \& Storey
(1982).  Lines emitted from the same upper levels depend only on
relative transition probabilities; this is the case for the
$\lambda$1399.77 and $\lambda$1407.39 lines where a ratio of 1 was
calculated, the $\lambda$1397.20 and $\lambda$1404.81 lines where a
ratio of 0.13  was calculated. The density dependent components
of \oiv were taken from Nussbaumer \& Storey (1982).  The emission
line features are consistent with  a density of $n_H=10^{10}\,\rm cm^{-3}$
given a temperature of the gas of $1.5\times 10^5 K$.   We point out
however that this feature is very weak and blended, and we cannot use
it to constrain the density of the emitting gas.

Fig.\ 3 demonstrates that our
assumption that the line profile is uniformly Lorentzian is well
justified, in most  cases.  There are a few discrepancies.   As
discussed in Section 2.1, the \ion{O}{6} line is near the edge of the
Lif1a detector, so that the only the 1032\AA\/  component is measured.   
We also found that the blue component is weaker than
the red component of the \ion{Mg}{2} doublet. This is not expected
physically, because the statistical weight of the blue component is
twice that of the red component.  We investigated the possibility that the
profile was affected by bad diodes or Galactic absorption lines, but
were unable to find the reason for this discrepancy. Also, we found
that the \heii line displays a blue wing; this was also reported by
Puchnarewicz et al.\ (1998). This is the only line in the
spectrum that displays such a feature.  It is possible that it
originates in recombination in optically thin, highly ionized gas.

In Fig.\ 5, we plot the velocity width of the brighter
($F_\lambda>1.5\times 10^{-14}\rm\, erg\, cm^{-2}\, s^{-1}$ for the
multiplet), and least-blended emission lines as a function of their
ionization potential (\ion{O}{6}, Ly$\alpha$, \ion{N}{5}, \ion{C}{4},
\ion{Si}{3}], \ion{C}{3}], \ion{Mg}{2}).  We find that most of the 
higher-ionization lines have slightly larger velocity widths
than  lower-ionization lines.  Such trends have been seen before
(e.g., Baldwin 1997), and they are consistent with our understanding
based on reverberation mapping results (e.g., Peterson 1998)
that the broad-line region is typically radially stratified.  The
exception is \ion{He}{2}. This line is different than the other
high-ionization lines in that it has a blue wing; it is possible that
deblending the blue wing artificially narrowed the line core.
RE~1034$+$39 is unusual in that the line profiles have the same shape
and are all centered at their rest wavelength; in other objects, high
ionization lines are not only broader but sometimes strongly
blueshifted (e.g., Leighly  \&  Moore 2004a).  RE~1034$+$39 is also
unusual in that the range of velocity widths of lines of different
ionization is small.   

The measured equivalent width of the 1031.9\AA\ component of the
\ion{O}{6}  line is 29\AA\/.  As noted in \S 2.1, we could not measure
the 1037.6 \AA\ component because of high background.  If we assume
that the gas is very optically thick, these two emission lines will
have the same equivalent width, and the equivalent width of the
doublet would be 58\AA\/.  Alternatively, in the optically thin
regime, the 1037.6 \AA\/ component should have half the flux of the
1031.9\AA\/ component, and the total equivalent width would be
43\AA\/.  In either case, given the low luminosity of
$\log{L_{2500}}=28.3$ [$\rm ergs\, s^{-1}$], we find that it is
roughly consistent with the \ion{O}{6} Baldwin effect presented by
Kriss (2001).  

In Table 4, we present a comparison of the bright emission line fluxes
with that of Ly$\alpha$ for RE~1034$+$39 and the same quantities
obtained from three composite spectra: the LBQS composite (Francis et
al.\ 1991), the radio-quiet composite from {\it HST} spectra (Zheng et
al.\ 1997), and  a composite spectrum from the FIRST Bright Quasar Survey
(Brotherton et al.\ 2001).  We find that the biggest difference is in
the measurement of \ion{O}{6}: the value from RE~1034$+$39 is much
larger with respect to Ly$\alpha$ than from any of the other
composites.  It should be noted, however, that the LBQS and FIRST
composites are made with observed-frame optical spectra, where 
Ly$\alpha$ forest absorption could reduce the flux of \ion{O}{6}.
However, this does not explain the difference with respect to the {\it
HST} composite.  RE~1034$+$39 also has stronger \ion{He}{2}
and weaker \ion{Mg}{2} than the composites.

\subsection{The Spectral Energy Distribution of RE 1034+39}

The spectral energy distribution (SED) of RE 1034+39 constructed from
the {\it HST}, {\it FUSE}, {\it EUVE} and {\it ASCA} data is shown in
Fig.\ 6. For comparison, we show the {\it Cloudy} AGN
continuum.  The strong soft excess peaking in the EUV first reported
by Puchnarewicz et al.\ (1995) is clearly seen.

The shape of the SED is frequently roughly parameterized by
$\alpha_{ox}$, the point-to-point power law slope between 2500 \AA\
and $\sim2 \rm \,keV$. We infer $\alpha_{ox}=1.22$ for RE~1034$+$39.
Wilkes et al.\ (1995) find that this parameter is correlated with the log
of the luminosity at 2500 \AA.  For an object with
$\log_{10}(l_{opt})=28.3$ (for $H_0=50$, $\Lambda_0=0$), their
regression predicts $\alpha_{ox}=1.29$.  It should be noted that 
there are only 5 objects with lower optical luminosity than
RE~1034$+$39 in the Wilkes et al.\ sample; these objects cluster
around $\log L_{2500} \approx 28$, and their $\alpha_{ox}$'s are $\sim
1.6$.  This suggests that RE~1034$+$39 has a markedly flat
$\alpha_{ox}$ compared with other AGN of its luminosity.

Puchnarewicz et al.\ (1995) reported that the optical/UV continuum could
be described by a power law \footnote{$\alpha_{opt}$ defined by
$F_{\nu} \propto \nu^{-\alpha_{opt}}$} with index $\alpha_{opt}\sim
0.9$.  This is somewhat steeper than that of the typical quasar (0.33
with rms dispersion of 0.59; Natali et al.\ 1998).  Our reanalysis
of the {\it HST} spectra, in combination with the {\it FUSE} spectrum,
shows that the spectrum subtly steepens at short wavelengths, and is 
consistent with a flatter slope of $\alpha_{opt}=0$ shortward of 2200
\AA\/ up to \ion{O}{6}.  The subtle upturn is seen in the inset in
Fig.\ 6.  Shortward of \ion{O}{6} the slope appears to become
flatter in the {\it FUSE} spectrum.  We have no explanation for this;
it is possible that the object is not in the slit for all detectors at
all times, creating a flux misalignment between the separate segments
of the spectrum. Because the continuum is not detected in the SiC1a
detector (Fig.\ 1), we cannot use that spectrum to check and repair
the flux alignment (e.g., Brotherton et al.\ 2001).  Regardless, we
observe, as did Puchnarewicz et al.\ 1995, that the optical/UV
spectrum does not extrapolate to the flux in the soft X-rays.

\section{One-zone Photoionization Models using the RE~1034$+$39
  Continuum} 

A primary goal of this paper is to see whether RE~1034+39's emission
line equivalent widths and ratios are consistent with with its unusual
spectral energy distribution.  To do this, we perform photoionization
modeling using the photoionization code {\it Cloudy} (Version 94.00;
Ferland 2001).   In this section, we first attempt to model the lines
with a one-zone model, making the simplifying assumption that the gas
emitting the lines is described by a single photon flux and density.
This assumption is partially justified by the similar profile and
velocity width of the emission lines (\S 2).  In \S 4, we relax this
assumption, and compute a Locally-Optimally emitting cloud model for
RE~1034$+$39.

{\it Cloudy} requires as input the spectral energy distribution, the
photon flux, the hydrogen number density, and the column density of
the gas.  To compute the line equivalent widths, a covering fraction
is required.  Gas metallicity can also be specified; we found
reasonably good results with solar metallicity, so we did not vary
this parameter. 

To obtain the spectral energy distribution, we used the observations
shown in Fig.\ 6.  The region between the {\it FUSE} spectrum, which
extends to 960\AA\/ in the rest frame, and the {\it EUVE} point at
0.156~keV is not observable, and we simply extrapolate using a power
law over this region.  Outside of the observed band pass, we used the
{\it Cloudy} AGN SED.  Specifically, we assumed that the spectrum
continues toward longer wavelengths to 10 microns before dropping
sharply toward the radio.  Also, as there is no spectral information
above 10 keV for this object, we extrapolate the hard X-ray power law
to 100 keV before requiring it to drop off sharply toward higher
energies.  Compared with the {\it Cloudy} AGN continuum, that of
RE~1034$+$39 begins to increase toward the UV at shorter wavelengths,
and is much stronger in the soft X-rays.

The photon flux, $\Phi$, expressed in $\rm photons\, cm^{-2}\,
s^{-1}$, and the hydrogen number density, $n_H$, expressed in
$cm^{-3}$, were the parameters that we sought to constrain with the
simulations.  To do so, we computed grids of models over a range of
$\Phi$ ($16.0<\log(\Phi)<20.75$; $\Delta\log(\Phi)=0.125$) and $n_H$
($8.5<\log(n_H)<11.5$; $\Delta\log(n_H)=0.125$).  Thus, we examined gas
conditions over a large range of ionization parameter
($-4.0<\log(U)<-0.5$).  

For the column density, we use the parameter $\log(N_H^{max})$, which is
defined as  $\log(N_H)-\log(U)$, where $N_H$ is the column density
(see also Leighly 2004). This parameter is a convenient one for studies of
photoionized gas because it adjusts the column density to account for
the ionization parameter. Therefore, for a particular value of
$N_H^{max}$, we probe to the same depth in terms of hydrogen 
ionization fraction regardless of the ionization parameter.
Initially, we tried a model that is optically thin to the hydrogen
continuum, in which  $\log(N_H^{max})$ was set equal to 21.5.  This
model was not able to produce the line equivalent widths and ratios;
this is not surprising, because observed low-ionization lines  such as
\ion{Mg}{2} are produced near the hydrogen ionization
front. Ultimately, we used $\log N_H^{max}=26$ because the slab is
then sufficiently thick to produce observed low-ionization lines such as
\ion{C}{2}.  Emission line fluxes and ratios for models that are
optically thin to the hydrogen continuum can be very sensitive to the
column density (e.g., Leighly 2004); optically-thick models are much
less sensitive, and therefore we fix this parameter.

The similar line profiles for the high-, intermediate-, and
low-ionization lines in this object suggest that all the lines are
produced in gas with similar or uniform properties.  Even though this
may not be a precisely accurate physical description of the emission
line region, as a first step, we investigate whether one region of the
$\Phi$ -- $n_H$ parameter space can produce all the lines.  To do
this, we plot the contours for which the model equivalent widths equal the
measured equivalent widths for all of the most prominent emission
lines (e.g., Baldwin et al.\ 1996).  The equivalent width  of a line
depends on the gas covering fraction.  We varied this parameter, and
determined that we could obtain a solution in which the most contours
overlapped in the smallest region for a covering fraction of 0.1.  The
results are shown in Fig.\ 7 for the  high-ionization lines
(\ion{O}{6}, \ion{N}{5}, \ion{C}{4}, and   \ion{He}{2}), and for the
intermediate- and low-ionization   lines (\ion{C}{3}~$\lambda 
977$, \ion{C}{2}~$\lambda   1335$, \ion{O}{3}]~$\lambda 1666$,
\ion{Si}{3}], \ion{C}{3}], and   \ion{Mg}{2}).    Ly$\alpha$ was not
plotted.  We found that the predicted value of Ly$\alpha$ equivalent
width was $\sim 1.9$--$3.1$ times higher than the observed value in
the box-enclosed region.  This may indicate that a slightly enhanced
metals abundance may be appropriate. 

We also plot the
 ratio of \ion{C}{3}]~$\lambda 1909$ to \ion{Si}{3}]~$\lambda
1892$.  This ratio is useful for constraining the density because
 these lines are produced under roughly the same physical conditions
 (e.g., Hamann et al.\ 2002), yet they have different critical
 densities ($3.4 \times  10^{9}\rm \, cm^{-3}$ and $1.04 \times
 10^{11}\rm\, cm^{-3}$).  As  seen in Fig.\ 7, there seems to be some
 dependence on  the flux (and therefore ionization parameter) as well.

For the high-ionization lines, seen in Fig.\ 7, we find that there
is a region in which the contours of all of the lines are close to one
another near $\log(\Phi)=18.5$, and $\log(n_H)=9.7$, which corresponds
to $\log(U)\approx -1.7$.  The contours show that the peak of the
\ion{C}{4} emission lies in this region, while \ion{O}{6}, \ion{N}{5}
and \ion{He}{2} peak at higher photon fluxes.  

For the intermediate- and low-ionization lines, seen in Fig.\ 7,
there is a region in which contours of all of the lines are close to
one another near $\log(\Phi)=18.2$, and $\log(n_H)=9.7$, which
corresponds to $\log(U)=-2$.  The contours show that the peak of the
\ion{C}{3}]~$\lambda 1909$ and \ion{C}{3}~$\lambda 977$ emission lies in
  this region, while \ion{C}{2}, \ion{O}{3}],
\ion{Si}{3}], and   \ion{Mg}{2} peak at slightly lower fluxes.  

The rectangles showing the location of the best solutions for
strongest high-ionization lines (Fig.\ 7) and the strongest
intermediate- and low-ionization lines overlap in a region in which
the ionization parameter lies in the range $-2.35 < \log(U) < -1.65$,
and in which the density is $\log(n_H)=9.75$.  Thus, we demonstrate
that the line emission in RE~1034$+$39 is consistent with emission
from gas with uniform properties, as is suggested by the similar line
profiles, and is consistent with illumination of that gas by the
observed extreme spectral energy distribution.

\section{Locally Optimally Emitting models}

AGN monitoring campaigns have shown that emission line fluxes vary in
response to continuum flux changes, with different lag times that
depend primarily on the ionization potential of the line.  This means
that the broad-line region is in general radially stratified.  Baldwin
et al.\ (1995) proposed that perhaps the quasar emission line region
is filled with clumps of gas of differing densities and distances from
the source of the ionizing continuum.  If this were the case, emission
lines would naturally be produced over an extended region, peaking
where the conditions for their emission are optimal.  This is known
as the Locally Optimally Emitting Cloud model, or LOC model.  We note
that there is evidence in some cases that the emission 
region is truncated in extent, ionization, or column density.  For
example, double-peaked broad line radio galaxies require truncation at
the outer edge of the disk to create double peaks (Eracleous \&
Halpern 2004).  In addition, the lack of intermediate- and
low-ionization line emission in the broad blueshifted lines in some
NLS1s required truncation of the column density in the wind component
(Leighly 2004).  Nevertheless, the LOC presents an arguably more physically
realistic scenario than the one-zone model presented above.

In this section, we present our second photoionization analysis:  an
LOC model for RE~1034$+$39.  At the 
same time, we recompute the LOC model using the parameters from
Baldwin et al.\ (1995) for comparison.  We first describe the
assumptions of the model, and our scheme for integration, before
describing the results.

\subsection{Setting up the LOC model}

Individual clouds are modeled with the same parameters used by Baldwin
et al.\ (1995).  Specifically, each single cloud has a hydrogen column
density of $N_H=10^{23} \rm \, cm^{-2}$, and is characterized by a
single value of hydrogen density $n_H$ and photon flux $\Phi$. In
order to reproduce Baldwin et al.\ results, we do not use 
$N_H^{max}$ for the column density, as described above. A single
column density for all clouds is a good assumption as long as all
clouds are optically thick to the continuum.  Following Balwin et al.,
we discuss the results in terms of line ratios, and therefore the
covering fraction does not enter into our calculations. We use solar
abundances.

We ran the LOC models for two spectral energy distributions.  We use
the one we inferred for RE~1034$+$39, described in \S 2.6.  For
comparison, we also compute models for the one used by Baldwin et al.\
(1995) that has a bump that peaks at 48 eV\footnote{In   the {\it
    Cloudy} parameter file, this model is  specified by ``AGN
  T=800000,-1.4,-0.3,-1.0''}.
We computed a grid of models for each continuum over the same range of
parameters used by Baldwin et al.\ (1995): density $8<\log n < 14$
with $\Delta \log n=0.125$ ($\rm cm^{-3}$); photon flux $18< \log \Phi
< 23$ with $\Delta \log \Phi=0.125$ ($\rm photons\,cm^{-2}\,s^{-1}$).

\subsection{Distribution Function and Integration}

We follow Baldwin et al.\ (1995) to construct our LOC models.  
The emission line strengths in LOC models are computed by first
assuming a function for the distribution of clouds in the
density/photon flux plane, then using this as  a weighting function
to integrate over the line emission obtained from the grid with the
following integral, which gives the luminosity of each line $L$:

\begin{equation}  L=\int
r^2 F_{line}(r,n)\Psi(r,n)drdn.
\end{equation}

The {\it Cloudy} models yield the line flux at each value of
$(n_H,\Phi)$.  The photon flux is related to the radius by $r \propto
\Phi^{-1/2}$, and  $F_{line}(r,n)$ is the line flux from the {\it
Cloudy} models.  $\Psi(r,n)$ is the distribution function.  It is
assumed to be a separable function of radius and density, so that
$\Psi(r,n)=f(r)g(n)$.  We use a power law for the distribution
functions, so that $f(r)=r^\alpha$ and $g(n)=n^\beta$.  Because the
{\it Cloudy} grids are a function of $\Phi$ rather than of $r$, it is
convenient to express $f(r)$ as a function of $\Phi$ instead.  Using
the assumption that $r\propto \Phi^{-1/2}$, $dr \propto \Phi^{-3/2}d\Phi$,
and $d(\log\Phi) \propto d\Phi/\Phi$, we find that $f(\Phi) \propto
\Phi^{-(\alpha+1)/2}$.

In order to obtain the integrated line emission, we use a
two-dimensional integration scheme employing the trapizoidal method
over the grid that is uniform in  
$d(\log n)$ and $d(\log \Phi)$.  This method requires us to determine a
weight for each grid point from the distribution functions.  The
result is $w_{ij}=C \times \Phi_i^{-(\alpha-1)/2}n_j^{\beta+1}$, where
  $C=1$ for the interior of the grid, is $1/2$ for the edges, and is
  $1/4$ for the corners. 

For each line, $F_{ij}$ is the line flux from the {\it Cloudy} model,
and $C_{ij}$ is the monochromatic continuum flux under Ly$\alpha$ at
every grid point $(i,j)$.  Thus, the value $F_{ij}/C_{ij}$ is
proportional to the equivalent width.  To compute the integral, we
 sum the weights over the grid as follows:

\begin{equation}
  L \propto \sum_i \sum_j {{F_{ij}}\over{C_{ij}}} w_{ij} d\log n d\log\Phi.
\end{equation}

Using these values of the line luminosity $L$ for each line we can
compute the ratio of each line with Ly$\alpha$.  

\subsection{LOC results}

The ratios of the line fluxes to Ly$\alpha$ are given in Table 5 for
both the RE~1034$+$39 and the {\it Cloudy} AGN continuum.  We adopted the
distribution functions used by Baldwin et al.\ (1995) in order to
compare explicitly with their results ($f(r)=r^\alpha$ with
$\alpha=-1$; $g(n)=n^\beta$ with $\beta=-1$).  Note that Baldwin et
al.\ (1995) erroneously report that they used $f(r)=constant$ and $g(r)
\propto n^{-1}$ (Korista 2003, p.\ comm.)

We first verify that we can reproduce the Baldwin et al.\ (1995)
results. The results taken directly from Baldwin et al.\ (1995) are
given in column 2, and our re-creation of those results is given in
column 3.  We find slight differences due to the difference in
versions of {\it Cloudy}; their computations were carried out with
Version 90.02d, while ours were done with Version 94.00.

In column 4 we give the results from the RE~1034$+$39 SED, and, for
comparison, the measured values from the spectra in column 5.  While
we are not able to match the line ratios exactly, the trends that we
find are encouraging.  That is, lines that are stronger with respect
to Ly$\alpha$ in the RE~1034$+$39 LOC model compared with the AGN
continuum LOC model are also stronger in the data.  These include
\ion{O}{6}+Ly$\beta$, \ion{N}{5}, \ion{O}{4}]+\ion{Si}{4}, \ion{C}{4},
and \ion{O}{3}]+\ion{He}{2}.  It is notable that the highest
ionization line that we measure, \ion{O}{6}, is much stronger in both
the data and the RE~1034$+$39 SED model than in the AGN LOC model.
This fact verifies our expectation that the extremely hard SED should
yield strong high ionization lines. Generally speaking, an extremely
hard SED should create stronger collisionally-excited  emission lines,
because each ionizing photon carries more energy on average. 

The largest differences between the ratios predicted by the LOC model
for RE~1034$+$39 and the measured ratios are for \ion{C}{4}, for which
the model predicts about twice the observed value, and in both cases,
the values are larger than those from the AGN continuum.  A very large
discrepancy is found for \ion{Mg}{2}, where the model predicts about
five times the observed value, and the observed value is less than
that predicted from the AGN continuum.  This discrepancy might have
been larger had we integrated to lower flux values in the LOC
computation, as \ion{Mg}{2} is optimally emitted at $\log(\Phi) \sim
17$ (e.g., Korista et al.\ 1997).  We have noticed that {\it Cloudy}
models seem to frequently overpredict \ion{Mg}{2} fluxes; a similar
over-prediction by about a factor of two was found in models for windy
NLS1s given in Leighly (2004).

In Fig.\ 8, we show the line emissivity as a function of photon flux,
assuming a 2-d geometry, for both the RE~1034$+$39 continuum, and the
AGN continuum.  In other words, we show the flux expected in each ring
of area $2 \pi r^2 dr$.  The emissivity has been normalized by the
luminosity in the entire line; i.e., we plot $L(R)/L_{total}$ for each
line.  For both continua, we see that significant line emission is
predicted for the LOC model for fluxes lower than $\log \Phi \le \sim
22$, out to at least $\log \Phi=18$.  However, the velocity widths of
the lines may not support such a broad emission region. Assuming
Keplerian velocities, a \ion{C}{4} velocity width of $900 \rm
\,km\,s^{-1}$ (Table 3), and an \ion{O}{6} velocity width of $1300 \rm
\,km\,s^{-1}$ corresponds to a ratio of fluxes of 4.4 at the radii
where these lines should dominate, or $\Delta \log \Phi \approx
0.6$. This is a much smaller difference than the difference in the
peaks of the \ion{C}{4} and \ion{N}{5} distribution of $\Delta \log
\Phi \approx 2.0$.  

It is also interesting to note that the distribution
of line emission is not much different for the two continua. The peaks
for the higher-ionization lines are at slightly larger radii for
RE~1034$+$39, except for \ion{C}{4}. The emissivity as a function of
photon flux for \ion{Mg}{2} is indistinguishable for the two SEDs.
Another interesting thing about this plot is that the intermediate-
and low-ionization lines increase strongly at larger radii (low
flux).  As discussed by Baldwin et al.\ (1995), the LOC model assumes
truncation of line emission at $\Phi < 18$ as a result of graphite
grains that will form in the clouds and decrease the emissivity.
However, the steep increase in emissivity at larger radii predict that
the observed flux of these intermediate- and low-ionization lines will
depend very sensitively on the precise radius at which dust starts to
dominate.  This sensitivity possibly conflicts with the observation
that the equivalent widths of these lines don't vary very much from
object to object.  

\section{Exploring the Influence of the Spectral Energy Distribution}

In \S 3 we presented photoionization models using the RE~1034$+$39
SED, and found that there is a region in the $\Phi$/$n$ plane that can
roughly reproduce the equivalent widths of the lines that we measured.
In \S 4, we presented an LOC model using the RE~1034$+$39 continuum,
and noted differences in predicted emission lines from the relatively
hard RE~1034$+$39 and the softer AGN continuum that correspond to our
expectation that a hard continuum should produce stronger
high-ionization lines, and a hard continuum should produce stronger
collisionally-excited emission lines. 

In this section, we present the third type of photoionization
analysis: we wish to determine whether the emission lines in 
RE~1034$+$39 are not only consistent with its observed hard SED, but
actually require it.  That is, is it possible to produce the observed
emission lines with a much softer SED, for some combination of
ionization parameter and density?   At the same time, we want to
examine the influence of the shape of the spectral energy distribution
on the emission lines in a more general way, for several reasons.
First, the only other systematic investigation of BLR emission-line
dependence on SED that we are aware of is Krolik \& Kallman (1988),
who consider only three distinct SEDs, and a handful of lines.  We
want to consider a larger range of SEDs and a larger number of lines.
Furthermore, we are also planning to use this in further
investigations of the influence of the SED on emission lines (in,
e.g., PHL~1811, an object with a relatively soft SED; Leighly et al.\
in prep.).  Finally, we use these results to investigate whether line
ratios currently commonly used as density and metallicity indicators
are sensitive to the SED.  However, so that the flow of our discussion
of the properties of RE~1034$+$39 is not interrupted, we relegate most
of this discussion to the Appendix (\S 7).  Briefly, we created a set
of semiempirical spectral energy distributions parameterized by the
cutoff temperature of the big blue bump $kT_{cut}$, which is measured
in $eV$. We then used  {\it Cloudy} (Version 96) to construct a grid
of model line fluxes for a range of densities and ionization
parameters.  We calculate {\it a posteriori} (see \S 7.4) the effect
of the global covering fraction. 

\subsection{Constraints using the RE~1034$+$39 Emission Lines}

Using the results presented in the Appendix, we can now address the
question: is the hard spectral energy distribution that we observed
from RE~1034$+$39 required in order to generate the observed
emission-line properties?  
In order to address this question, we follow a procedure similar to
that used in Leighly (2004) to constrain {\it Cloudy} models for
extreme NLS1s.  We compare the line measurements from the observed
spectra with the modeled line fluxes from the grids of {\it Cloudy}
models using a figure of merit ($FOM$),
defined as the absolute value of the difference of the base-10 log of
the {\it Cloudy} line fluxes, transformed to the value that would be
measured on Earth using the procedure described in the Appendix
\S{7.4},  and the observed line fluxes.  Note that
in Leighly (2004) the $FOM$ was defined in terms of the line
equivalent widths. Line fluxes work better here because the shape
of the continuum underneath the lines changes as the SED changes; in
Leighly (2004) that continuum was held constant.   Thus the $FOM$ is
defined as 
\begin{equation}
FOM_{jkml}=\sum_i
\vert\log(F^{model}_{here,ijkml})-\log(F^{measured}_{here,i})\vert
\end{equation}
where $i$, $j$, $k$, $m$, and $l$ index the individual lines, the
    $kT_{cut}$ describing the SED, the $\log(U)$, the $\log(n_H)$, and
    the covering fraction, respectively.

The $FOM$ is defined using \ion{O}{6}~$\lambda$1034,
\ion{N}{5}~$\lambda$1240, \ion{C}{4}~$\lambda$1549,
\ion{He}{2}~$\lambda$1640, 
\ion{C}{3}]~$\lambda$1909, and \ion{Si}{3}]~$\lambda$1892.  These
are the six  highest-equivalent-width lines measured in
RE~1034$+$39, excluding Ly$\alpha$, which we discuss below, and
the low-ionization line \ion{Mg}{2}.  As discussed in 
\S{7.3}, the semiforbidden lines \ion{C}{3}] and \ion{Si}{3}] are
commonly used as density indicators, and thus   allow us to
constrain the density in the line emitting gas in   RE~1034$+$39. 

We computed the $FOM$ for the entire grid of {\it Cloudy} models. The
minimum value of the $FOM$ located the spectral energy  distribution,
ionization parameter, density and covering fraction that yielded line
fluxes that matched most closely the observed fluxes.  In Leighly
(2004), evidence for  multiple minima was found, so here we looked for
multiple minima in a brute force way by sorting the data in order of
increasing $FOM$ and looking for jumps in the parameter values that
would indicate another minimum. We also explored the $FOM$ space by
perturbing various parameters and examining the contours.  After a
thorough search, we found no evidence for multiple minima.

The minimum $FOM$ was found for the following parameters:
$kT_{cut}=240\, \rm eV$, $\log(U)=-1.8$, $log(n_H)=9.125$ and a covering
fraction of $6\%$. In Fig.\ 8, we show covering fraction/density, and
$kT_{cut}$/$\log(U)$ cuts in $FOM$ space around this minimum.  It is
important first to notice that the minima are somewhat broad; as
discussed in Leighly (2004), broad minima are arguably physically more
realistic because they imply that even though we are using a one-zone
model, gas with properties perturbed a reasonable amount from the
properties of the minimum will produce nearly identical lines so that
the solution is not overly finely tuned.  However, the $FOM$ increases
strongly toward lower $kT_{cut}$ for values less than $\sim 150\rm \,
eV$, indicating a harder SED is strongly favored.  

Overlaid on Fig.\ 8 are the contours of the emission lines used to
construct the $FOM$.  We can see the confluence of the contours near
the $FOM$ minimum.  The grid of models described in the Appendix only
include $\log(U) \leq -1.125$, so we originally could not see if there
were any other minima at higher ionization parameters.  So we extend,
for the best fitting values of the covering fraction and density, the
$\log(U)$ vs $kT_{cut}$ plane to higher values of $\log(U)$.  Note
that there is little density dependence for $\log(n_H) < 11.0$ for the
high-ionization resonance lines, and, everything else being equal, the
covering fraction only influences the flux, so we lose nothing by not
considering those two parameters explicitly here.  This extension
reveals that the contours illustrating the measured values of the
highest ionization lines (\ion{O}{6}, \ion{N}{5}, and \ion{He}{2}) do
not dip below $kT_{cut}=110\rm \,eV$. This indicates that a high
$kT_{cut}$ SED is supported even at high ionization parameter, and
that perhaps the SED/$\log(U)$ degeneracy is broken.

Our model, however, is not without limitations.  In Fig.\ 9 we show
the log of the ratio of the observed data to the model at the $FOM$
minimum, both for the strong lines used to generate the $FOM$, and
Ly$\alpha$, \ion{Mg}{2}, \ion{Si}{4} $\lambda 1397$, and \ion{C}{3}
$\lambda 977$ as well.  A prominent difference is seen for Ly$\alpha$.
One possible explanation for this difference is that the gas in
RE~1034$+$39 has somewhat enhanced metallicity; our {\it Cloudy}
models consider only solar metallicity gas.  We note that we
found no similar discrepancy for the LOC models.  We also find prominent
discrepancy for \ion{Mg}{2}, such that the model value is much too
strong.  We find this also for the LOC models for RE~1034$+$39, as
discussed in \S{4.3}, and we have no explanation for this.
\ion{Si}{4} appears to be somewhat too strong; this is a weaker line
that is generally blended with \ion{O}{4}]~$\lambda$ 1404, and it may
be that we prescribe too large fraction of the 1400\AA\/ feature to
\ion{Si}{4}.  

In addition, this is a one-zone model; it assumes that all the lines
are produced by gas with uniform ionization parameter, density and
covering fraction.   As discussed earlier, this assumption may be
fairly good for RE~1034$+$39 in particular, since the high-
and intermediate-line profiles are very similar.  But comparison of
Fig.\ 8 and Fig.\ 9 for the highest ionization lines may indicate the
weakness in this assumption.  Specifically, at the best $FOM$ point,
we produce too much \ion{C}{4} and insufficient \ion{O}{6},
resulting in discrepant observed-to-model ratios in Fig.\ 9.  This 
discrepancy may be alleviated in a stratified model if a somewhat
higher ionization parameter for the high-ionization lines is used.
Nevertheless, the fact that the contours for the high-ionization lines
lie at high $kT_{cut}$ regardless of the ionization parameter may
indicate that even for a stratified model, a relatively hard SED is
still supported.

\section{Discussion}

\subsection{Summary of Results}

We have presented spectra and analysis from simultaneous {\it FUSE}, {\it
ASCA}, and {\it EUVE} observations of the Narrow-line Seyfert 1
Galaxy RE~1034$+$39.  Reanalysis of archival {\it HST} spectra were also
presented.  We find that almost all the lines in RE~1034$+$39 can
be described well by narrow ($\rm FWHM \leq \sim 1200 \rm \,
km\,s^{-1}$) Lorentzian profiles centered at the rest  wavelength.
That this is true for the UV emission lines was previously reported
from the {\it HST} spectra by Puchnarewicz et al.\ (1998); our
{\it FUSE} spectrum reveals for the first time the strong, narrow
\ion{O}{6} line.  The line profiles in RE~1034$+$39 are different
than those in some other Narrow-line Seyfert 1 galaxies, in
which only the intermediate- and low-ionization lines are narrow and
centered at the rest wavelength; the high-ionization lines are broad
and strongly blueshifted (e.g., Leighly \& Moore 2004).  
Our simultaneous far-UV and X-ray observations reveal a spectral
energy distribution that peaks in soft X-rays, as was previously found
by Puchnarewicz et al.\ (1995).  This spectral energy distribution is
much harder than that from typical quasars.  

We presented three complementary types of photoionization analysis.
First, we use the observed RE~1034$+$39 spectral energy distribution,
and assume that because the emission lines all have very similar
profiles, they all may be emitted by gas with approximately uniform
ionization parameter and density.  The presence of low-ionization
emission lines implies that the gas is optically thick to the
continuum.  The ratio of the semiforbidden lines \ion{C}{3}] and
\ion{Si}{3}] constrains the density of the gas to be around
$\log(n_H)=9.75$.  The equivalent widths of the other strong
emission lines are consistent with a photon flux of $\log(\Phi)
\approx 18.4$, implying an ionization parameter near
$\log(U)=-1.8$.  

As has been discussed by Baldwin et al.\ (1995; see also Korista et
al.\ 1997), the line-emitting gas may not be characterized by a single
ionization parameter and a single density but rather there may be a
distribution of gas properties, with the emission of any particular
region being dominated by emission from lines most favorably produced
there; this is the Locally Optimally-emitting Cloud or LOC model.  The
second photoionization analysis presented is an LOC model for
RE~1034$+$39.  We computed the ratios of the emission lines with
Ly$\alpha$ for LOC models using the RE~1034$+$39
SED and gas and density distributions used previously by Baldwin et
al.\ (1995).  For comparison, we reproduce the Baldwin et al.\
results.  We find that the LOC results correspond quite well with
the observations from  RE~1034$+$39. In particular, \ion{O}{6} is
strong, as observed.  An exception is \ion{Mg}{2}, which is observed
to be about 5 times weaker than predicted by the model.  

Finally, we investigated the role of the spectral energy distribution
on the emission line fluxes explicitly by constructing a set of
semiempirical spectral energy distributions parameterized by the
cutoff  temperature of the blue bump, and computing a large grid of
line fluxes as a function of density, ionization parameter, and
spectral energy distribution.  Following Leighly (2004), we define a
figure of merit ($FOM$) as the sum of the absolute value of the
difference between the model and observed fluxes for six strong
emission lines.  We find a fairly broad minimum in the FOM around
$kT_{cut}=240 \rm \,eV$, $\log(U)=-1.8$, $\log(n_H)=9.125$, and covering
fraction of 6\%.  The RE~1034$+$39 continuum most closely matches a
semi-empirical SED with $kT_{cut} \sim 230\rm \,eV$, so it is not
surprising that the minimum FOM solution has nearly the same
ionization parameter and density as was inferred in the
photoionization analysis using the observed RE~1034$+$39 continuum.
Furthermore, the $FOM$ increases strongly toward lower values of
$kT_{cut}$ corresponding to softer continua, indicating that a hard
continuum, in the context of a one-zone photoionization model, is
necessary to produce the emission lines that we see; softer continua
simply do not produce the strong high-ionization lines that are the
notable feature of the RE~1034$+$39 spectrum.

\subsection{The \ion{O}{6} Line in RE~1034+39}

In this section, we compare the \ion{O}{6} observed in RE~1034$+$39
with that observed in other AGN.  RE~1034$+$39 is a somewhat
exceptional object compared with most of those objects that have
\ion{O}{6} measurements in that it has a relatively low luminosity,
and a hard spectral energy distribution characterized by a flat
$\alpha_{ox}$. Such a hard spectral energy distribution is conducive
for strong \ion{O}{6} emission; does RE~1034$+$39 match this expectation?  

Zheng, Kriss \& Davidsen (1995; hereafter ZKD) find a significant
anticorrelation between the \ion{O}{6}/Ly$\alpha$ ratio and
$\alpha_{ox}$ in a sample of quasars observed on the ground or with
{\it HST}, {\it IUE}, or {\it HUT}.  They present a regression for
this anticorrelation: $\log($\ion{O}{6}/Ly$\alpha)=-0.74 \alpha_{ox}
+0.38$.  Based on our measured $\alpha_{ox}=1.22$, this regression
predicts that \ion{O}{6}/Ly$\alpha$ will be 0.3.

We only measure the 1032\AA\/ component of \ion{O}{6}.  Under
optically-thin conditions, the ratio of the 1032\AA\/ component to the
1037\AA\/ component is 2.0.  Under very optically-thick conditions, that
ratio would be 1.0.  Using the values for the line fluxes in Table 3,
this analysis indicates a \ion{O}{6}/Ly$\alpha$ ratio between 0.76 and
1.0.  So, the \ion{O}{6} flux relative to Ly$\alpha$ appears to be
particularly large compared with other objects, and it can be seen
from their Fig.\ 1 that this ratio would have been the largest in the
ZKD sample.  However, the regression of ZKD has uncertainties
representing the spread in values; taking into account the
uncertainties shows that the observed ratio  can be explained.
In addition, there are some differences in the way the data were
analyzed.  We detect \ion{O}{5}]~$\lambda 1218$, and our model removes
that contribution from the Ly$\alpha$ flux.  That was not done in the
Zheng et al.\ sample, making their values of Ly$\alpha$
slightly larger, and hence their values of \ion{O}{6}/Ly$\alpha$
slightly smaller.  However, if we combine our measured fluxes for
Ly$\alpha$ and \ion{O}{5}, it changes our ratio range to 0.59--0.78,
still large compared with the rest of the ZKD sample.

Zheng, Kriss \& Davidsen (1995) also compare the equivalent width of
the objects in their sample with the UV monochromatic luminosity.
Using their cosmology ($H_0=50\rm \,km\,s^{-1}\, Mpc^{-1}$; $q_0=1$)
we find that $\log(L_{1550})=28.1$ [$\rm erg\,s^{-1}$] for
RE~1034$+$39.  This value makes 
RE~1034$+$39 less luminous than all but one of the objects considered
by Zheng et al.  Using their regression for the anticorrelation
between the \ion{O}{6} EW and the luminosity, we find a predicted
\ion{O}{6} equivalent width of 150\AA\/.  The observed equivalent
width of \ion{O}{6}~$\lambda$1032 component is 29\AA\/, implying the
equivalent width of the whole feature is between 43 and 58 \AA\/.  So,
the predicted equivalent width is higher by greater than a factor of
two.  Considering the regression is between logarithmic quantities, the
discrepancy is probably acceptable. Furthermore, our observed
equivalent widths are consistent with the range of equivalent widths
implied by the uncertainties on the regression parameters.

We conclude that, although RE~1034$+$39 has extreme \ion{O}{6}
properties, they are consistent with trends appropriate for an object
with its observed spectral energy distribution and luminosity.  This
suggests that neither the emission line region geometry nor the
inclination angle is especially unusual in this object, but rather is
an extension of those in other quasars.

\subsection{Directly Comparable Previous Results}

Krolik \& Kallman (1988) were possibly the first to address the issue
of the response of various emission lines to the shape of the spectral
energy distribution, with the aim of trying to determine the shape of
the continuum in the unobservable extreme ultraviolet.  They consider
three spectral energy distributions that are different in the shape of 
the extreme UV.  They compute emission line ratios for combinations of
three ionization parameters and three pressures, finding that some
ratios vary negligibly between the different SEDs, while some
ratios show considerably greater range.  They attribute these results
to some extent to the relative strength in the continuum bands
responsible for exciting production of particular emission lines.  For
example, \ion{He}{2}~$\lambda 1640$ and \ion{O}{6} are sensitive to the
helium continuum, and therefore tend to be stronger with respect to
Ly$\alpha$ for SEDs with strong helium continuum compared with Lyman
continuum. 

In the Appendix, we present similar computations as were presented by
Krolik \& Kallman 
(1988).  A difference is that we examine a much larger range of
spectral energy distributions and ionization parameters, although we
do limit ourselves to discussion of a single density and slabs that
are optically thick to the continuum in this paper.  Another
difference is that we 
examine a larger number of lines.  Our inferences are more or less the
same for our harder continua (\S 7.2).  However, our results are
somewhat different for the very soft continua that were not
investigated by Krolik \& Kallman (1988).  For these continua, the
low-ionization lines, which are produced below the hydrogen ionization
front for the harder continua, are produced above the hydrogen
ionization front and become primary coolants.  While not directly
applicable in this paper, the results of low-kT simulations are
important for understanding objects with soft continua such as
PHL~1811 (Leighly et al.\ in prep.)

\acknowledgements

We thank Jules Halpern for reducing the {\it EUVE} data, and we thank
Chiho Matsumoto for reducing the {\it ASCA} data.  DC acknowledges
helpful discussions with John Moore, and with Kirk Korista regarding
the LOC models. We thank Joe Shields for suggesting the identification
of \ion{O}{5}~$\lambda 1218$.  We acknowledge useful comments from an
anonymous referee that resulted in substantial improvement in the
paper.  Some of the data presented in this paper were obtained from
the Multimission Archive at the Space Telescope Science Institute
(MAST).  STScI is operated by the Association of Universities for
Research in Astronomy, Inc., under NASA contract NAS5-26555.  Support
for MAST for non-HST data is provided by the NASA Office of Space
Science via grant NAG5-7584 and by other grants and contracts.  This
research has made use of the NASA/IPAC Extragalactic Database (NED)
which is operated by the Jet Propulsion Laboratory, California
Institute of Technology, under contract with the National Aeronautics
and Space Administration. This research has made use of data obtained
from the High Energy Astrophysics Science Archive Research Center
(HEASARC), provided by NASA's Goddard Space Flight Center.  KML and
DAC gratefully acknowledge support by NASA grants NAG5-10171 (LTSA),
NAG5-10038 (FUSE), and NAG5-10366 (FUSE). EB gratefully acknowledges
partial support by NASA grants NAG5-3505 and NAG5-12127, NSF grant
AST0204771, and computer resources at NERSC supported by the US DOE.

\section{Appendix}

\subsection{Photoionization Modeling with a Semi-Empirical SED}

To test the influence of the spectral energy distribution on the line
emission, we needed a series of SEDs that could be expressed simply
mathematically, and yet are approximately physically realistic.  We
start with the AGN spectrum that is included in the {\it Cloudy}
package:

\begin{equation}
    f_{\nu}=\nu^{-\alpha_{uv}}e^{\frac{h \nu}{k T_{cut}}}e^{\frac{k
    T_{IR}}{h \nu}} + a \nu^{\alpha_x}.
\end{equation}

This equation models the big blue bump as a power law with index
$\alpha_{uv}$ in the optical and UV with an exponential cut off at
high energies parameterized by the temperature $T_{cut}$, and an
exponential cutoff at low energies parameterized by $T_{IR}$. A power
law with index $\alpha_{x}$ models the X-ray spectrum.  We fixed the
optical/UV power law index to $\alpha_{uv}=0.33$, adopting the
observationally-determined mean value from a sample of quasars by
Natali et al.\ (1998), which is in agreement with the mean obtained
from the LBQS by Francis et al.\ (1991), and with the Einstein sample
from Elvis et al.\ (1994).  We fixed the X-ray power law continuum
slope to $\alpha_{x}=1$ (e.g., Zdziarski, Poutanen, \& Johnson 2000).

For the thin disk model, the high energy cutoff for the optical-UV
continuum is theoretically related to the temperature of the inner
edge of the accretion disk.  For a fixed accretion rate with respect
to the Eddington value, and for a fixed efficiency of conversion of
accretion energy to radiation, we expect the accretion disk luminosity
to be proportional to $T^{-4}$.   The relative normalization between
the X-ray and UV is observed to be dependent on luminosity through
$\alpha_{ox}$, the point-to-point slope between 2500\AA\/ and 2 keV
(Wilkes et al.\ 1994).  We use the theoretical constraint provided by
the inner edge temperature, and the empirical constraint provided by
the $\alpha_{ox}$ to relate $T_{cut}$ to the X-ray power law
normalization $a$ through their mutual dependence on luminosity.  The
proportionality constant was determined using the measured $T_{cut}$
from coordinated {\it IUE} and {\it ROSAT} observations of the quasar
3C~273 (Walter et al.\ 1994).  Our final constraint relates the lower
limit frequency of the X-ray power law to $T_{cut}$ so that the power
law cannot be discerned in the optical-UV band.  

In the discussion that follows, the spectral energy distribution is
uniquely parameterized by the value of $kT_{cut}$.  We computed
photoionization models for a range of $kT_{cut}$ from $10\rm \,eV$ to
$320\,\rm eV$, with $\Delta kT_{cut}=10$.  A representative subset of the
spectral energy distributions, normalized so that they all have the
same ionizing flux, are shown in Fig.\ 11.  As shown in Fig.\ 12, a
spectral energy distribution with $kT_{cut}=230\rm\,eV$ coincides
fairly well with the spectral energy distribution of RE~1034$+$39, and
one with $kT_{cut}=10 \rm \, eV$ corresponds well with that of
PHL~1811, a luminous NLS1 (Leighly et al.\ 2001; Leighly, Halpern \&
Jenkins 2004; Leighly et al.\ in prep.).  This demonstrates that these
semiempirical SEDs are representative of observed SEDs.

Our primary goal is to determine whether the emission lines in
RE~1034$+$39 require a hard spectral energy distribution, or whether
they are independent of SED.  Thus, we
choose the remaining parameters for the photonization models to be
appropriate for RE~1034$+$39.  In \S 3 we found that the column
density should be large enough to be optically thick to the continuum.
Therefore, here we use $N_H^{max}=25$, which is sufficiently thick to
produce low-ionization lines such as \ion{C}{2}.  We verified that,
for a fixed ionization parameter, the depth of the hydrogen ionization
front is almost independent of the shape of the spectral energy
distribution (Fig.\ 13).  However, the depth of the helium ionization
front changes, being deeper for harder SEDs (Fig.\ 13).  This trend is
expected, as a larger proportion of the ionizing photons are in the
helium continuum for harder SEDs.

We run {\it Cloudy} models for SEDs with $10<kT<320$; $\Delta(kT)=10$,
for hydrogen densities $7.875<\log(n_H)<12.75$;
$\Delta(\log(n_H))=0.125$, and for ionization parameters $-3.0 <
\log(U)<-1.25$; $\Delta(\log(U))=0.125$.  Thus, 19,200 models were
considered.  

Ly$\alpha$ is a recombination line, and its intensity will depend
primarily on the depth of the hydrogen ionization front for the fixed
density considered in these models.  Thus, its intensity should depend
{\it only} on the ionizing flux, which is proportional to the ionization
parameter for a fixed density, and be independent of the shape of the
continuum, as long as the density is not so high that it starts to
become thermalized.  The contours of Ly$\alpha$ flux as a function of
ionization parameter and $kT_{cut}$ show that this is indeed the case
(Fig.\ 14).  Therefore, to partially remove the effect of the
ionization parameter, we present the results as ratios of lines with
respect to Ly$\alpha$.

\subsection{Model Results}

In this discussion, we consider only the dependence on the spectral
energy distribution and ionization parameter, and consider only one
density, $n_H=10^{10}\rm \, cm^{-2}$.  
Fig.\ 15--18 shows contours of emission lines with respect to
Ly$\alpha$ for the adopted range of $kT_{cut}$ and ionization
parameter.  We divide the results into four categories based on the
ionization potential of the emitting ion and the emission mechanism.
Figures 15--17 show the predominately collisionally excited lines,
with Fig.\ 15 showing lines with ionization potential (I.P.) greater
than that of \ion{He}{2} ($\rm I.P.> 54.4 eV$), Fig.\ 16 
showing lines with I.P. between that of hydrogen and helium ($\rm
13.6<I.P.<54.4$), and Fig.\ 17 with I.P. less than that of hydrogen
($\rm I.P.<13.6 \rm \, eV$).  Fig.\ 18 shows recombination lines
from hydrogen and helium. 

The behavior of some lines is quite easy to understand.  For
example, Fig.\ 15 shows the emission lines from ions with
ionization potential higher than that of He$^{2+}$.  They are emitted
most strongly when the continuum is relatively hard, because the
ionization potential of the ions that create these lines is relatively high.
They are emitted strongly when the ionization parameter is high
because a high ionization parameter is required for the high
ionization state.  In addition, the harder continuum generates a higher
temperature in the gas.  The lines shown here are those from
transitions with low excitation potentials, and are thus primary gas
coolants.  Therefore, the contours shown in Fig.\ 15 have maxima
along the top of the plot, where the continuum is hard, and tail
toward the right side, where the ionization is high.

Turning to Fig.\ 16, we find emission lines emitted between the
helium ionization front and the hydrogen ionization front.  In
particular,  \ion{C}{4} and \ion{N}{4}] are emitted most favorably at
intermediate values of the ionization parameter, near $\log(U)=-1.5$.
Again, these peak for the hardest continua because they are important
gas coolants.  

The situation is somewhat different for the intermediate ionization
lines \ion{O}{3}], \ion{Si}{4}, \ion{N}{3}~$\lambda 991$,
\ion{N}{3}]~$\lambda 1750$, \ion{C}{3}~$\lambda 977$, and
\ion{C}{3}]~$\lambda 1909$, shown in Fig.\ 16a.  These lines are most
favorably produced at somewhat lower ionization than the lines
discussed above.  Therefore, their contours peak along the top of the
plots at slightly lower ionization parameter (e.g., $\log(U) \approx
-2$ for \ion{C}{3}).  They also show enhancements along the right side
of the plots for continua with intermediate hardness.  This occurs
because in this region of the plots, the continua are too soft to
create abundant ions with highest ionization; intermediate ionization
states are dominant above the hydrogen ionization front.  This is a
consequence of the reduction in depth from the illuminated side of the
slab to the helium ionization front, and the corresponding increase in
depth of the intermediate ionization region between the helium and
hydrogen ionization fronts as the spectra become softer.  Thus, at
high ionization, the gas cools via production of intermediate
ionization lines.  The excitation of the ions is predominately
collisional in both regions of the plots, so permitted and
semiforbidden lines have roughly the same behavior.

The situation is radically different for the intermediate ionization
lines \ion{Al}{3} and \ion{Si}{3}]. The contours from these lines,
shown in Fig.\ 16b, peak at intermediate ionization parameter, but are
stronger for the softer spectra.  It is particularly interesting to
note the differing behavior for \ion{C}{3}] and \ion{Si}{3}]; the
contours shown in Fig.\ 16b are almost mutually exclusive.  The ions
responsible for these lines are isoelectronic.  Both are transitions
from the ground state to levels with the same configuration ($^3P^0$),
and their excitation potentials are quite similar.  The principle
difference between them appears to be the ionization potential to
create the $+2$ and $+3$ ions.  To create Si$^{2+}$ and C$^{2+}$, the
ions emitting these lines, 16.4 and 24.3 eV are required,
respectively.  To ionize them to the $+3$ state, 33.4 and 47.9 eV are
required, respectively.  Thus, Si$^{2+}$ is more favorably produced
when the continua are soft; for harder continua, the predominate
ionization state in the intermediate ionization region is Si$^{3+}$.
Furthermore, the {\it Cloudy} output indicates that for $\log(U)
\approx -2$ and for the softer spectra, \ion{Si}{3}] is the
predominant coolant.  \ion{Al}{3} behaves similarly, although it is a
permitted line, as the ionization potentials to create Al$^{2+}$ and
Al$^{3+}$ are 18.8 and 28.4 eV, respectively.

Moving to the low-ionization lines, we show contours from
\ion{N}{2}, \ion{C}{2}, \ion{Si}{2}, \ion{Al}{2}, \ion{Mg}{2},
\ion{Ca}{2}, and \ion{Na}{1} in Fig.\ 16b and 17.  The contours from these
lines exhibit  different behavior.  First we consider the pairs
of lines of \ion{N}{2}~$\lambda 1085$, \ion{N}{2}]~$\lambda 2140$, and
\ion{C}{2}~$\lambda 1335$, \ion{C}{2}]~$\lambda 2326$.  In both cases,
the semiforbidden line is most strongly emitted at low ionization
parameter for a hard spectrum, while the permitted line is most
strongly emitted at a high ionization parameter for a soft spectrum.
Although these are low-ionization lines, they are emitted in the
partially ionized zone only for the hard spectra and high ionization
parameter.  Examination of the {\it Cloudy} output shows that this
contrasting behavior originates in different excitation of the ion in
the two regimes.  At low ionization parameter, the semiforbidden line
is collisionally excited. For a particular ionization parameter, it
therefore emits most strongly for the harder spectra because the gas
temperature is higher.  The permitted lines are weaker in this regime
because they have much higher excitation potentials than the
semiforbidden lines, and for the low ionization parameters at which
the appropriate ions exist, the temperature is not high enough to
excite them.  They are emitted more favorably for soft spectra at high
ionization parameter as a result of recombination.

The behavior of the low-ionization line \ion{Si}{2} is interesting.
The Si$^+$ ion has five  groups of permitted resonance transitions
with upper level energies less than 11 eV, near 1160, 1263, 1308, 1531
and 1814 \AA\/.   For harder continua, the {\it Cloudy} output
indicates that the excitation is predominately collisional. Thus, the
1814 \AA\/ line is generally predicted to be the strongest of the
\ion{Si}{2} resonance lines, and is further predicted to be
stronger relative to Ly$\alpha$ at the lowest ionization parameters.  But
for the softer spectra, continuum pumping is the predominate process
for the higher excitation transitions.  In fact, for the  highest
ionization parameter and softest spectrum, the higher-ionization
transitions are as strong or stronger than \ion{Si}{2}~$\lambda 1814$.
This is an interesting result, because while \ion{SI}{2}~$\lambda 1263$
and \ion{Si}{2}~$\lambda 1308$ are strong in some AGNs,
\ion{Si}{2}~$\lambda 1814$ is not correspondingly strong, a result that
has been difficult to explain (Baldwin et al.\ 1996).  This issue will
be examined further in Leighly et al.\ (2004 in prep).

We turn next to the low ionization lines from Mg$^+$, Al$^+$, Ca$^+$, and
neutral sodium (Fig.\ 17).  These contours tend to be enhanced for the
hardest and softest spectra.  For the hardest spectra, these emission
lines are produced to a large extent in the partially ionized zone
beyond the hydrogen ionization front. The partially ionized zone is
predominately powered by X-rays.  As shown in Fig.\ 13, as the spectra
become softer, the depth of the partially-ionized zone shrinks,
resulting in relatively less emission at a given ionization parameter
for spectra with intermediate hardness.  For the softest spectra,
these lines are again produced strongly, because the ions emitting
these lines have low ionization potentials; thus these lines serve
relatively larger role as cooling agents when the spectra are soft.

Finally, we consider the recombination lines from hydrogen and helium.
The \ion{He}{2} lines are relatively easy to understand.  They are
strong when the helium ionization front is deep.  As seen in Fig.\ 13,
this is true when the continuum is hard.  In contrast, \ion{He}{1} is
strong when the region between the helium and hydrogen ionization
fronts is large; as seen in Fig.\ 13, this  condition is met when
the continuum is relatively soft.

\subsection{SED Dependence of the Density Diagnostic \ion{Si}{3}]/\ion{C}{3}]}

The ratio of the semiforbidden lines \ion{Si}{3}] and \ion{C}{3}] is 
frequently used to estimate the density of the emitting gas. These
emission lines are produced by isoelectronic ions and have similar but
not identical excitation potentials, and
therefore in principle should be produced under roughly the same
conditions.  The important difference is in their critical densities:
$1.04 \times 10^{11}\rm \,cm^{-3}$ and $3.4 \times 10^{9}\rm cm^{-3}$
for \ion{Si}{3}] and \ion{C}{3}] respectively.

The analysis presented above shows, however, that these emission lines
are not independent of the spectral energy distribution, as a
consequence of their different roles in cooling the gas, which may be
the consequence of the differing ionization potentials.  As discussed
above, for intermediate and soft spectra with ionization parameter of
$\log(U)=-2.5$, \ion{Si}{3}] is emitted strongly because it is one of
the primary coolants for the gas; for harder spectra, the ions
responsible for cooling are instead C$^{3+}$ and Si$^{3+}$.  Thus, a high
\ion{Si}{3}] to \ion{C}{3}] ratio could either indicate high density,
or a soft spectral energy distribution.  This is illustrated in Fig.\
19, which shows, for example, a range of greater than 2 in this ratio
{\it for a fixed density and ionization parameter} near
$\log(U)=-2.3$, as the SED varies from hard to soft.  Thus, there is a
degeneracy between density and spectral energy distribution for the
\ion{Si}{3}]/\ion{C}{3}] ratio that in principle could hamper its use
as a density indicator.

This degeneracy may be partially broken by using \ion{Al}{3} in
combination with \ion{Si}{3}] and \ion{C}{3}].  \ion{Al}{3} is also
predicted to be strong when the spectrum is soft, but it has little
dependence on density because it is a permitted line. Thus, if both
\ion{Al}{3} and \ion{Si}{3}] are strong with respect to \ion{C}{3}], a
soft spectrum may be indicated; if \ion{Al}{3} is weak while
\ion{Si}{3}] is strong, a high density may be indicated.

\subsection{SED Dependence of Metallicity indicators}

We looked at the SED dependence for three different metallicity
indicators (e.g., Hamann et al.\ 2002): \ion{N}{3}~$\lambda
1750$/\ion{O}{3}]~$\lambda 1666$, \ion{N}{5}~$\lambda
1240$/\ion{C}{4}~$\lambda 1550$, and \ion{N}{5}/(\ion{O}{6}+\ion{C}{4})
(Fig.\ 20).  Fig.\ 20a shows that \ion{N}{3}]/\ion{O}{3}] 
    has little 
dependence on the spectral energy distribution, and is therefore
arguably the most reliable metallicity indicator.  In contrast, Fig.\
20b shows that the \ion{N}{5}/\ion{C}{4} ratio has a strong dependence
on ionization parameter, and some dependence on SED for the softest
spectral energy distributions.  The ionization parameter dependence is
partially mitigated by using instead the
\ion{N}{5}/(\ion{C}{4}+\ion{O}{6}) ratio (Hamann et al.\ 2002) as
shown in Fig.\ 20c, except for the softest SEDs.

\subsection{Obtaining the Observed Model Line Fluxes for Comparison
  With Measured Values} 

For reference, in this section, we describe the method that we used to
compare the {\it Cloudy} results with the observed line fluxes.

The luminosity of the ionizing light observed here on Earth equals the
luminosity of the ionizing light at the continuum source, i.e., 

\begin{equation}
L_{here} =L_{there}.
\end{equation}

Therefore, if  $d$ equals the distance from the ionizing continuum
source to the observer, $\Phi$ equals the number flux of ionizing
photons from the continuum source, and $r$ equal to the distance from
the ionizing continuum source to the photoionized gas, we find that 

\begin{equation}
\Phi_{here} d^2 = \Phi_{there} r^2,
\end{equation}

so that 

\begin{equation}
\frac{\Phi_{here}}{\Phi_{there}}=\frac{r^2}{d^2}.
\end{equation}

If the clouds do not cover the ionizing continuum source completely,
the covering fraction $CF$ must be taken explicitly into account so
that 

\begin{equation}
r^2 CF F_{there}=d^2 F_{here}.   
\end{equation}

Eliminating the distances using Equation 7, and taking the base-ten
log, we obtain

\begin{equation}
\log(F_{here})=\log(\Phi_{here})-\log(\Phi_{there})+\log(CF)+\log(F_{there}).
\end{equation}

The figure of merit is defined as the absolute value of the difference
in the base-ten log of the  line fluxes, and is given by

\begin{equation}
FOM_{ijkml}=\vert \log(\Phi_{here})-\log(\Phi_{there,k})+
\log(CF_l)+\log(F^{model}_{there,ijkm})-\log(F^{measured}_{here,i})\vert 
\end{equation}

where $i, j, k$, and $m$ index the individual lines, the value of
$kT_{cut}$, the value of $\log(U)$, and the value of $\log(n_H)$,
respectively.

Assuming arbitrarily that $\log(\Phi_{there})$ is $17.75$, then we can
express $\log(\Phi_{here})$ as
\begin{equation}
\log(\Phi_{here})=17.75-\Delta \log(\Phi), 
\end{equation}

where $\Delta \log(\Phi)$ is the scaling factor required to reduce
the observed continuum with photoionizing flux of 17.75 to the
photoionizing flux we observe on Earth.  Comparing the {\it Cloudy}
output with the observed spectral energy distribution indicates that
$\Delta \log(\Phi)=18.25$.   Inserting Equation 11 in to the $FOM$
equation results in
\begin{equation}
FOM_{ijkml}=\vert 17.75-18.25-\log(\Phi_{there,j})+\log(CF_l)+
\log(F^{model}_{there,ijkm})-\log(F^{measured}_{here,i})\vert.
\end{equation}

\clearpage

\begin{figure}
\figurenum{1}
\epsscale{1.0}
\plotone{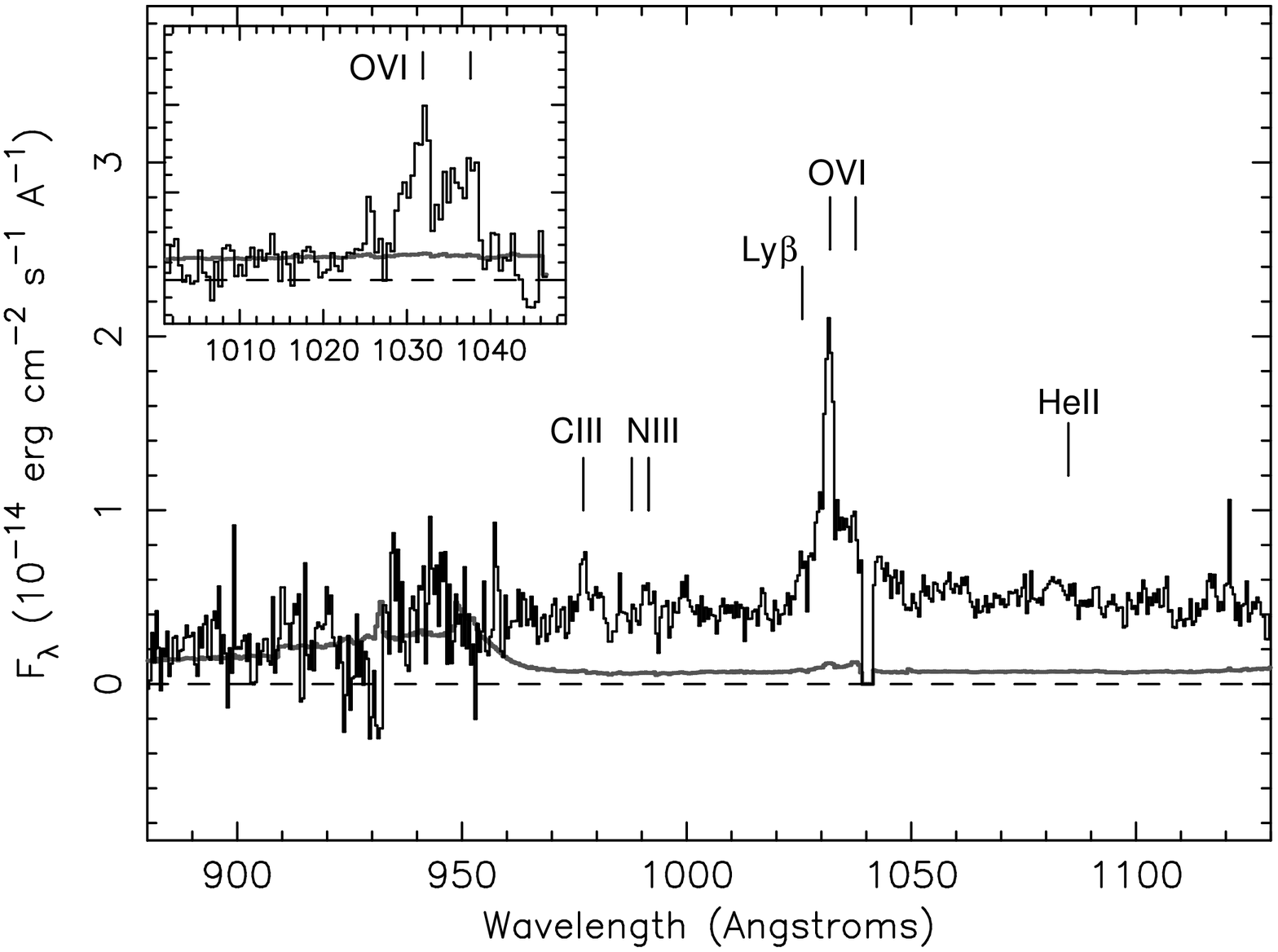}
\caption{\label{fig1} The extracted, coadded {\it FUSE} spectrum of
  RE~1034$+$39. 
  Airglow lines from Ly$\beta$, Ly$\delta$, and   Ly$\gamma$ have been
  removed. The thick line shows the uncertainty spectrum.  Emission
  lines commonly observed   in this bandpass in AGN are labeled; we
  detect \ion{C}{3},   \ion{O}{6}, and Ly$\beta$.  Note that the 1037
  \AA\/   component of the \ion{O}{6}   doublet was not detected in
  the LiF1a spectrum, as it falls on the edge of the detector where
  background is enhanced (see text for details).  Inset: the SiC1a
  spectrum, showing enhancements at the appropriate wavelengths for
  both components of the \ion{O}{6} line.}
\end{figure}

\clearpage

\begin{figure}
\figurenum{2}
\epsscale{1.0}\plotone{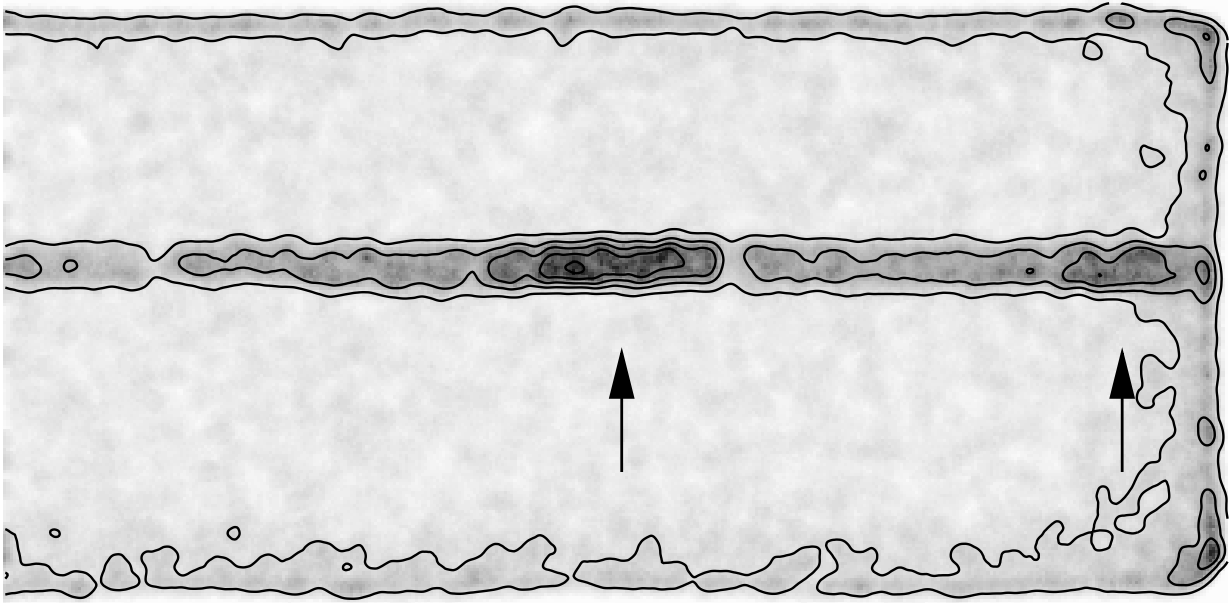}
\caption{The long wavelength portion of the LIF1 detector image, after
  screening and  before background subtraction. The image has been adaptively
  smoothed and is plotted on a linear scale.   The LIF1a trace is
  clearly detected.  The arrows show approximate expected centers of
  the \ion{O}{6} doublets.  The 1031.93\AA\/ component is clearly
  seen.  The 1037.62\AA\/ component can be seen near the edge of the
  detector, where the effective area is lower and the background is
  higher.}
\end{figure}

\clearpage

\begin{figure}
\figurenum{3}
\epsscale{0.95} \plotone{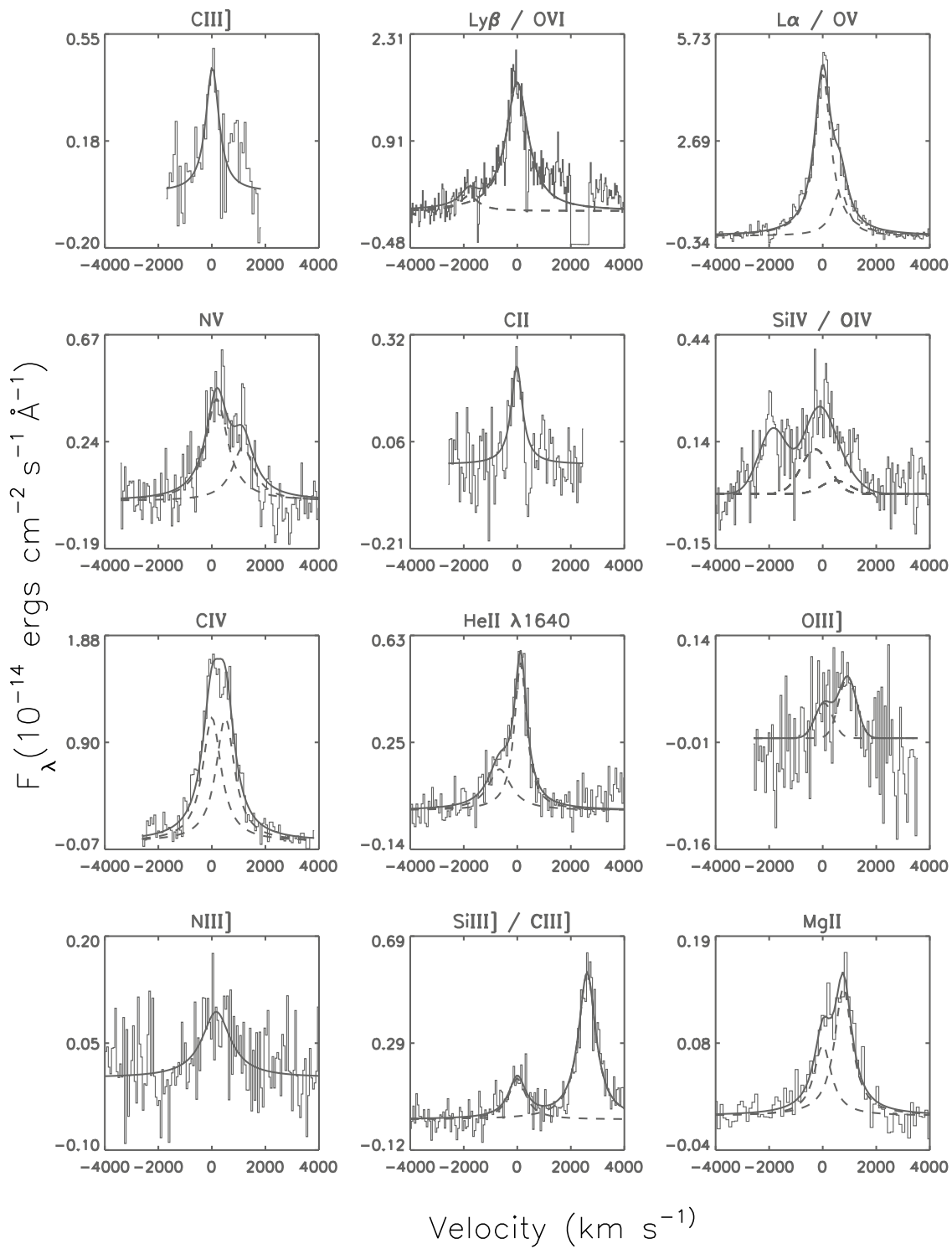}
\caption{The models for the emission lines in the
  RE~1034$+$39 spectra, after continuum subtraction.  All emission
  lines are consistent with a Lorentzian profile.}
\end{figure}

\clearpage

\begin{figure}
\figurenum{4}
\epsscale{1.0}\plotone{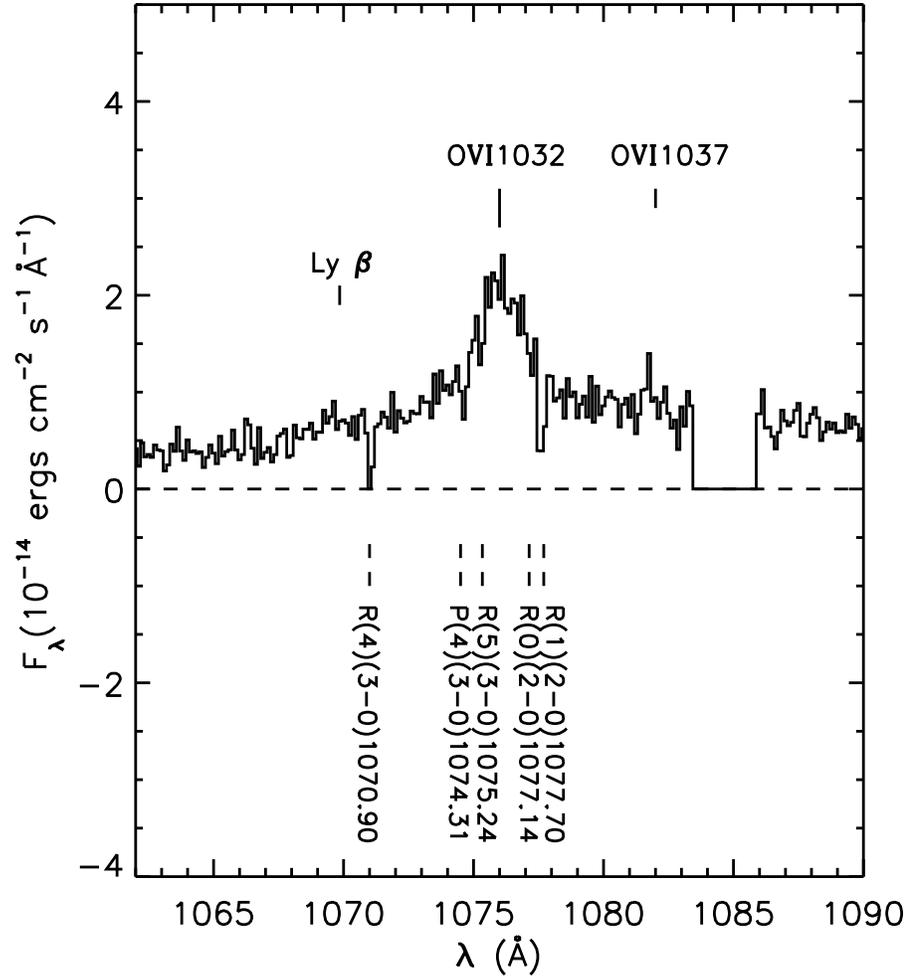}
\caption{\label{FUSEDATA2} The region of the spectrum containing
  \ion{O}{6} and Ly$\beta$. The rest-frame positions of the emission
  lines are labeled above the plot; the Galactic absorption lines are
  labeled below.  All absorption lines are consistent with zero
  redshift molecular hydrogen transitions originating in our Galaxy; no
  absorption lines intrinsic to the AGN are observed.}
\end{figure}

\clearpage

\begin{figure}
\figurenum{5}
\epsscale{1.0} \plotone{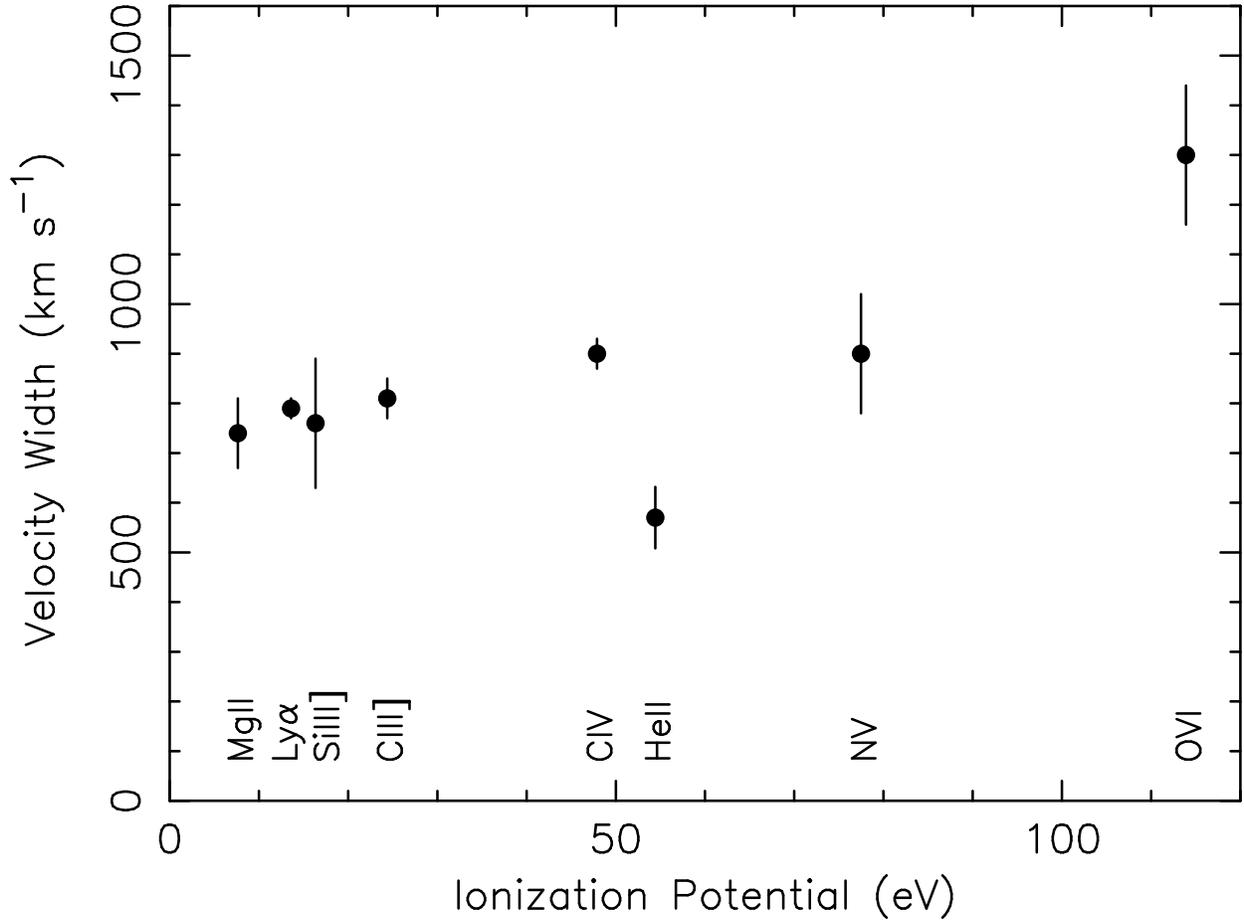}
\caption{The velocity width (FWHM) as a function of ionization
  potential for the brighter and least-blended FUV and UV emission
  lines.  A slight increase in velocity width potentially indicates
  stratification of the emission line region.  The exception is
  \ion{He}{2} which differs from the other lines because it has a blue
  wing; deblending the blue wing may result in a narrower core.}
\end{figure}
\clearpage

\begin{figure}
\figurenum{6}
\epsscale{1.0} \plotone{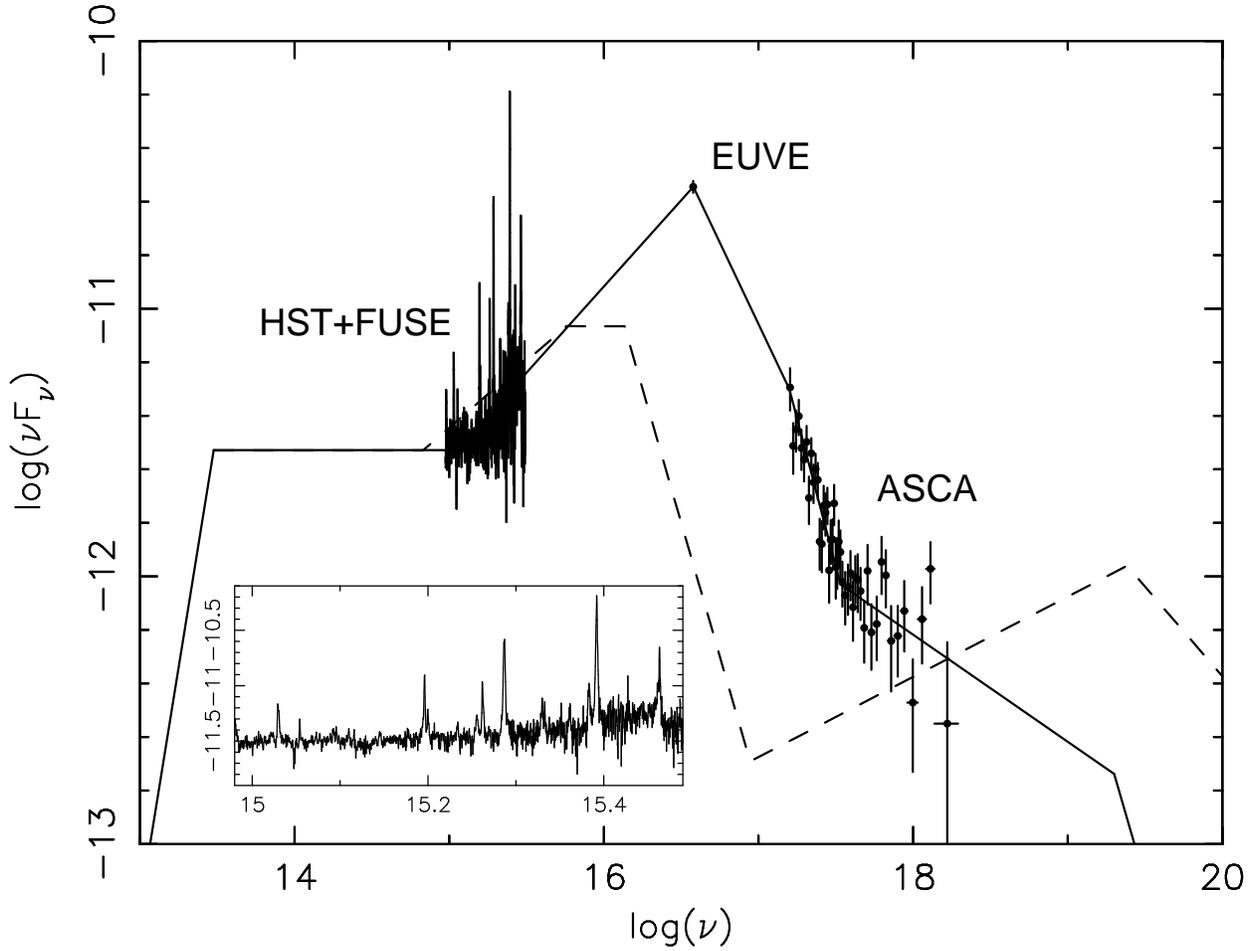}
\caption{\label{SED} The adopted continuum for {\it Cloudy}
  simulations was constructed from the simultaneous {\it FUSE}, {\it
  EUVE} and {\it ASCA} data, and the nonsimultaneous {\it HST} data.  The
  {\it Cloudy} AGN continuum is shown for comparison (dashed line).
  The inset shows the subtle upturn in the UV continuum toward high
  frequencies.}
\end{figure}

\clearpage

\begin{figure}
\figurenum{7}
\epsscale{1.0} \plotone{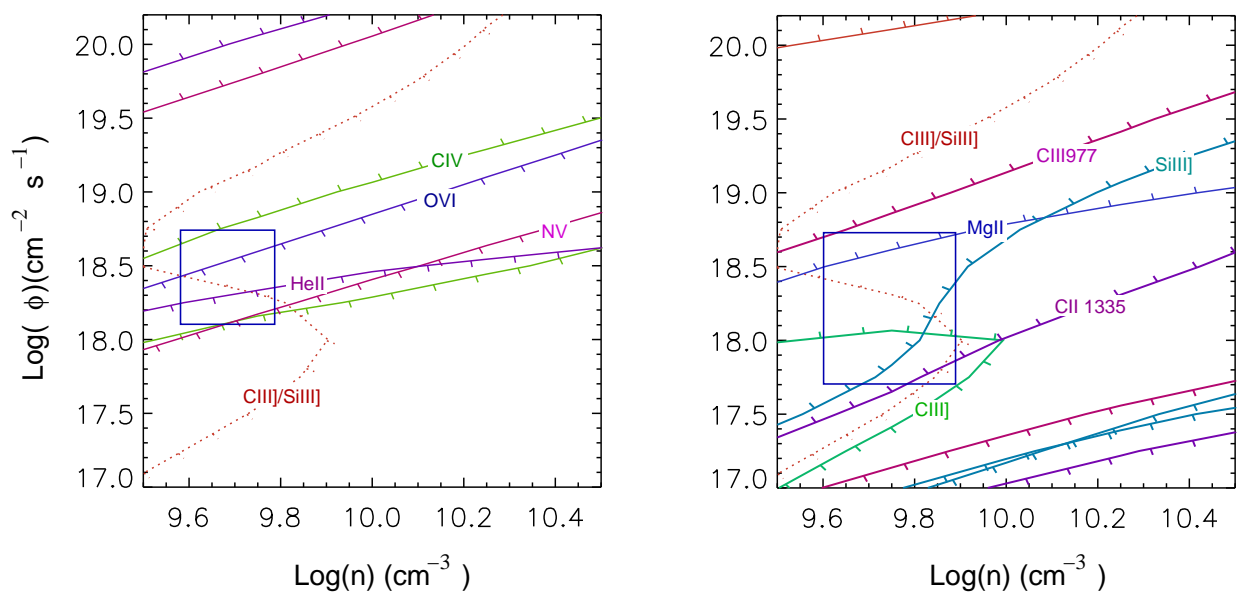}
\caption{Cloudy model results for the
 high-ionization lines (left) and the intermediate- and
  low-ionization lines (right).  The dotted line shows the ratio of
 \ion{C}{3}] to \ion{Si}{3}], a parameter that should be sensitive to
  density.  The contour ticks mark the direction of decreasing
 equivalent width. The boxes delineate the region of parameter space
 that is   intersected by   all contours.  }
\end{figure}

\clearpage

\begin{figure}
\figurenum{8}
\epsscale{1.0}\plotone{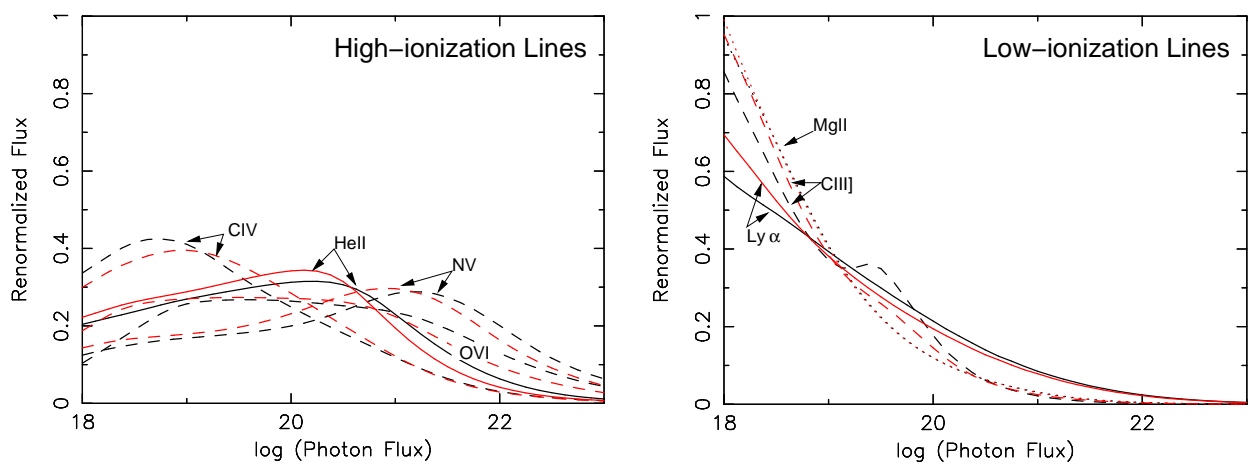}
\caption{Renormalized line emissivity for a 2-d, flattened geometry
  for LOC models. The left plot shows the results for the
  high-ionization lines, and the right plot shows the results for the
  intermediate- and low-ionization lines.  The red and black lines
  show the results for the  RE~1034$+$39 and AGN continua,
  respectively. Little difference is seen between these two continua;
  however, the emission is distributed over a much larger $\Delta \Phi$
  than is implied by the difference in the velocity widths of the lines.
    The emissivity of the low-ionization lines rises strongly toward
  large radii (low fluxes) perhaps indicating sensitivity of these
  lines to the value of the lowest $\Phi$ used in LOC models.}
\end{figure}

\clearpage

\begin{figure}
 \figurenum{9}
\epsscale{1.0} \plotone{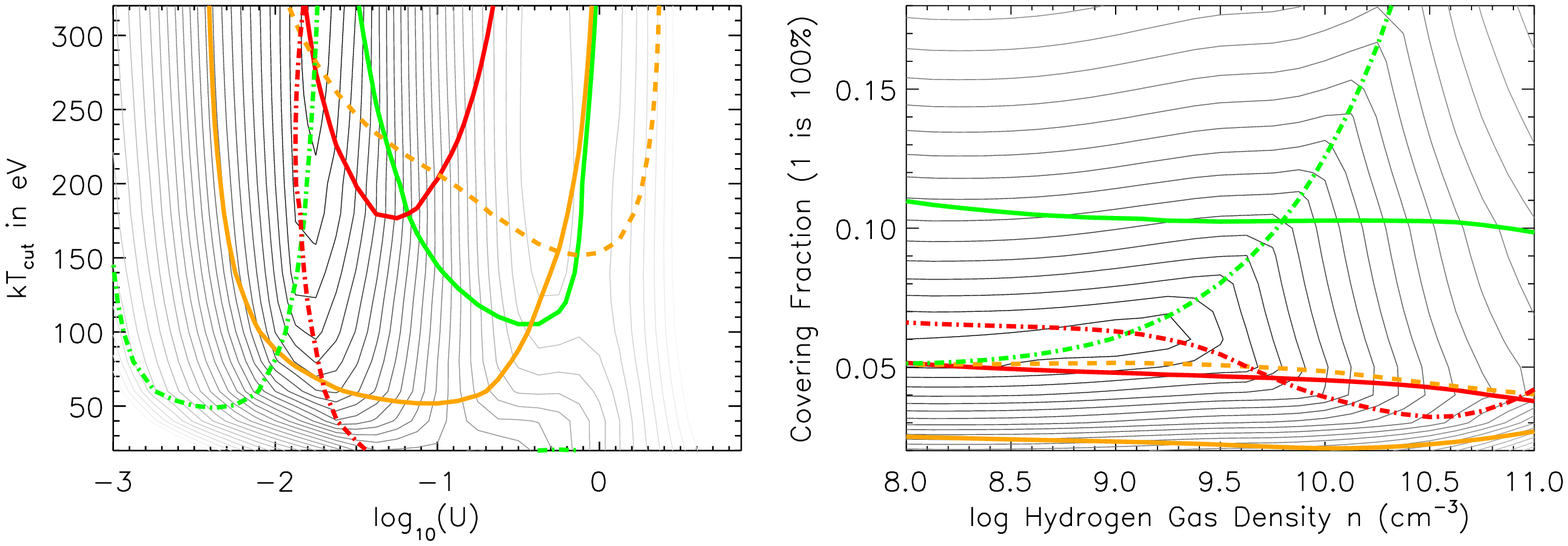}
\caption{Cloudy model results for cuts through the $FOM$ matrix around
 the global minimum located at $kT_{cut}=240\rm \, eV$, $\log(U)=-1.8$,
 $\log(n_H)=9.125$, and covering fraction of 6\%. The spectral energy
 distribution parameterized by 
 $kT_{cut}$ versus the ionization parameter $\log(U)$ ({\it left}) and the
 covering fraction versus the hydrogen density $\log(n_H)$ ({\it right}) are
 shown.  The thin, grayscale contours show the $FOM$ with interval of
 0.4,  and with the darker contours in the minimum.  The thick colored
 contours show where the model line fluxes equal the measured line
 fluxes with the following key: green -- \ion{O}{6}; red --
 \ion{N}{5}; orange -- \ion{C}{4}; dashed orange -- \ion{He}{2};
 dash-dot red -- \ion{Si}{3}]; dash-dot green -- \ion{C}{3}] (i.e.,
 permitted lines are solid, recombination lines are dashed, and 
intercombination lines are dash-dotted.) }
\end{figure}

\clearpage

\begin{figure}
\figurenum{10}
\epsscale{1.0} \plotone{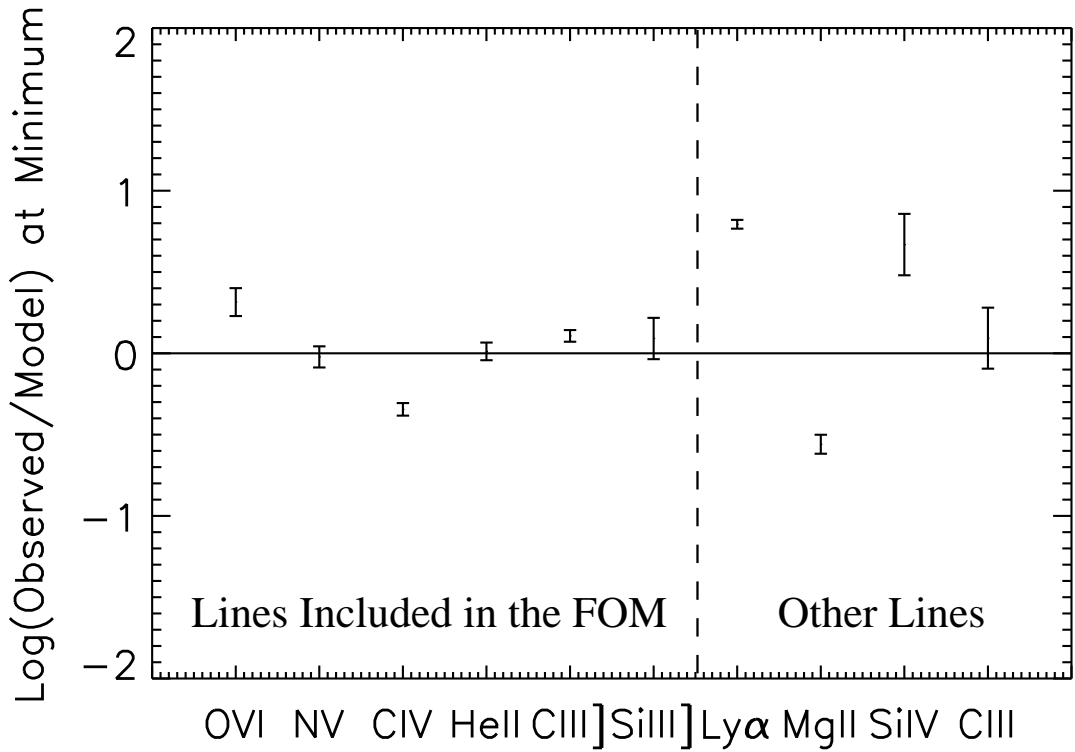}
\caption{\label{Tmodels} The log of the ratio of the observed line
fluxes with model line fluxes.  Error bars indicate statistical
uncertainty from the spectral fitting.  The six lines on the left were used to
construct the $FOM$. }
\end{figure}

\clearpage

\begin{figure}
\figurenum{11}
\epsscale{1.0} \plotone{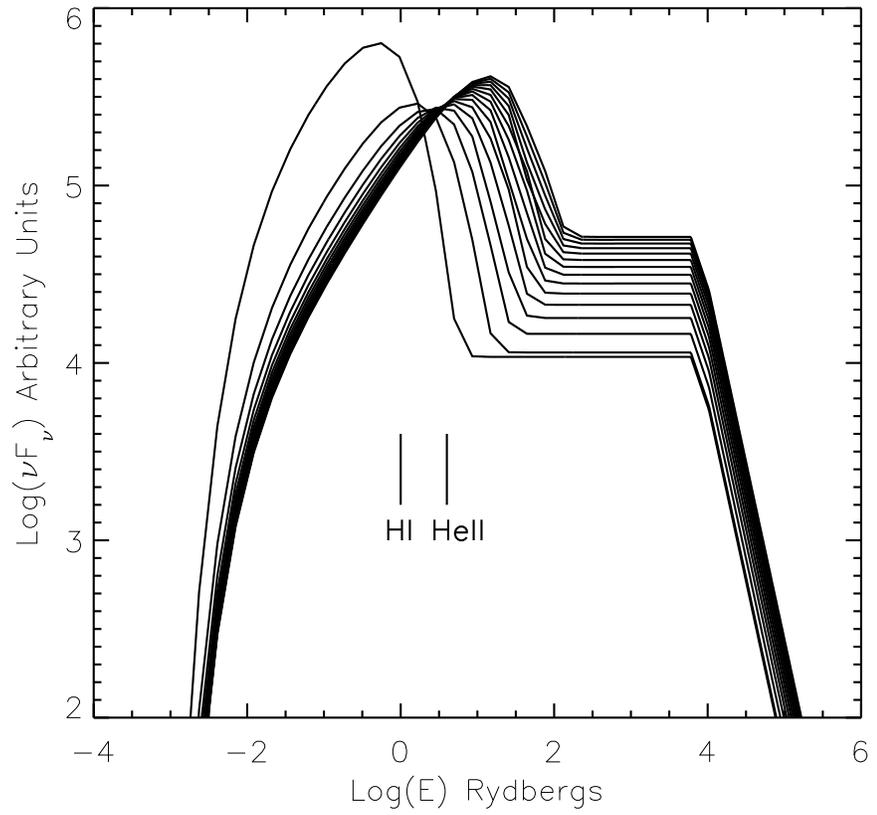}
\caption{A representative set of the semi-empirical spectral energy
  distributions used in \S 7, normalized so   that they all yield the
  same  value of the ionizing flux.} 
\end{figure}

\clearpage

\begin{figure}
\figurenum{12}
\epsscale{1.0} \plotone{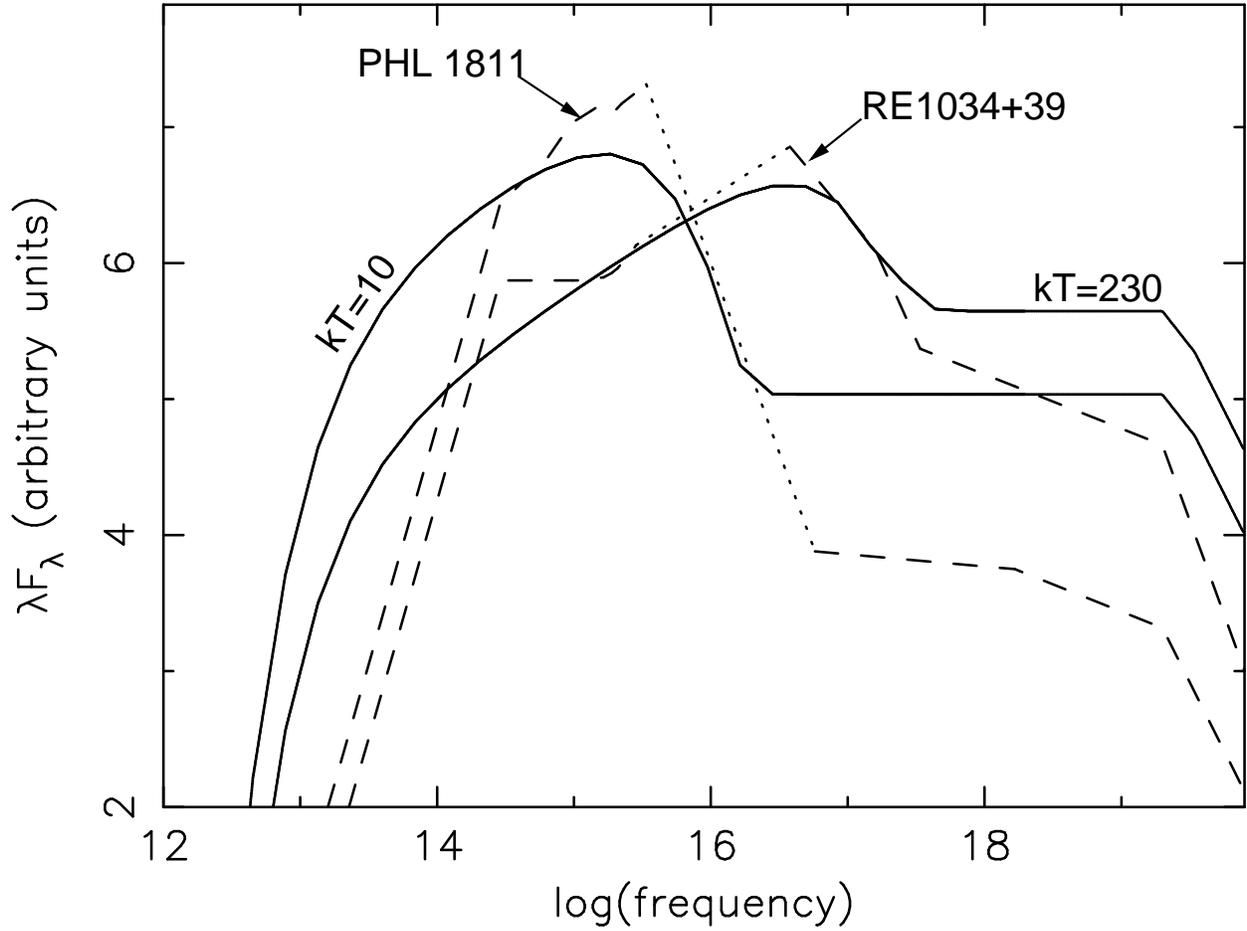}
\caption{Comparison of observed SEDs (solid lines) from coordinated observations
  with the semi-empirical spectral energy distributions (dashed lines;
  dotted lines in the unobservable EUV), illustrating that the
  semi-empirical SEDs correspond fairly well with observed ones.} 
\end{figure}

\clearpage

\begin{figure}
\figurenum{13}
\epsscale{0.8} \plotone{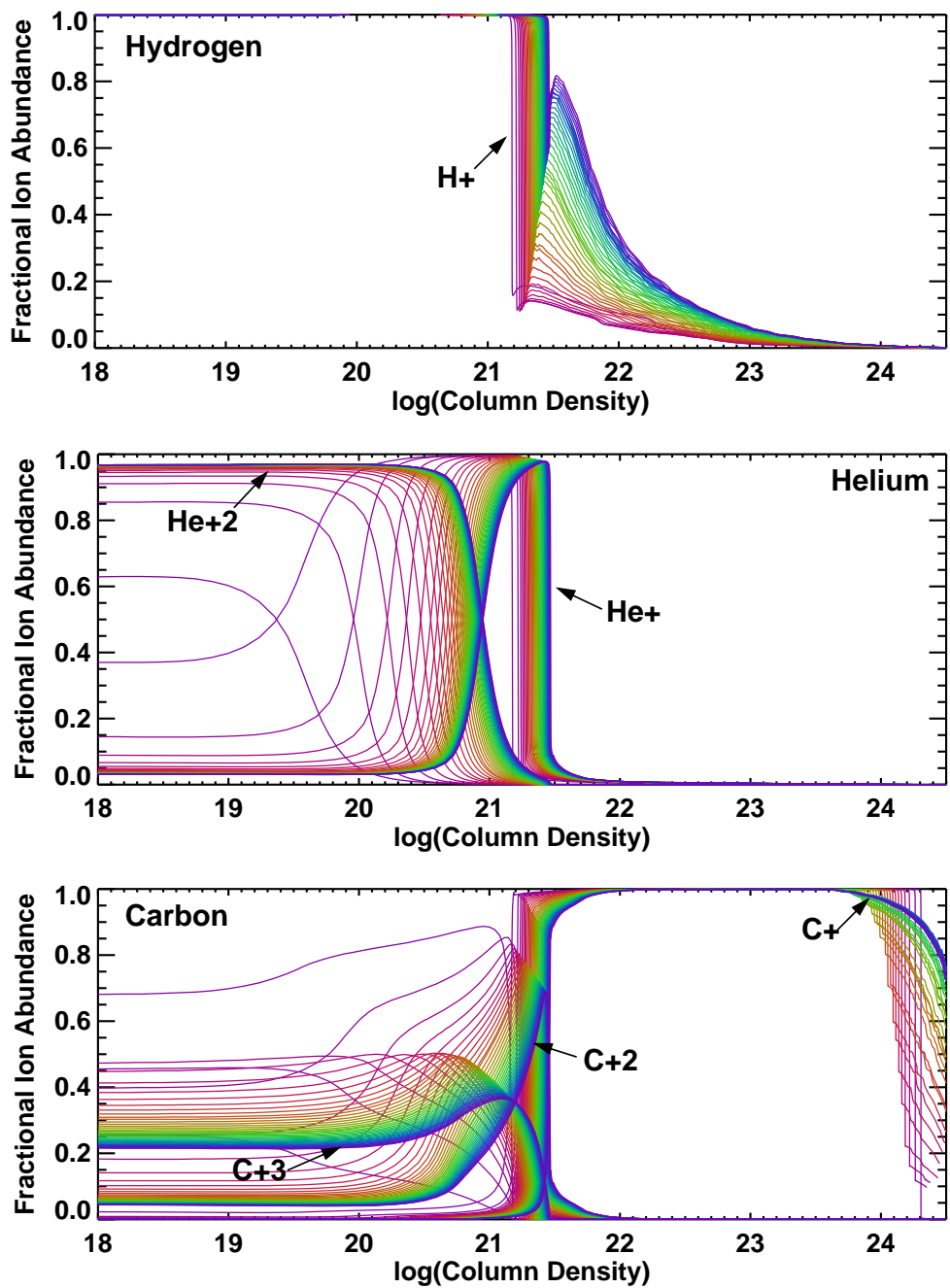}
\caption{The ionization fraction for select ions of hydrogen, helium
  and carbon as a function of column density for the semi-empirical
  SEDs.  We assume $\log U=-2$, $\log n_H=10$, and $\log N_H=26$ for
  illustration.  Red lines   corresponds to lower values of $kT$ (soft
  continua), and    blue corresponds to higher values of $kT$ (harder
  continua).   Notable differences include a deeper partially-ionized zone for
  hydrogen for the harder continua, deeper \ion{He}{1} region but
  shallower \ion{He}{2} region for the softer continua, and overall
  low ionization at small column density for softer continua.} 
\end{figure}

\clearpage

\begin{figure}
\figurenum{14}
\epsscale{1.0} \plotone{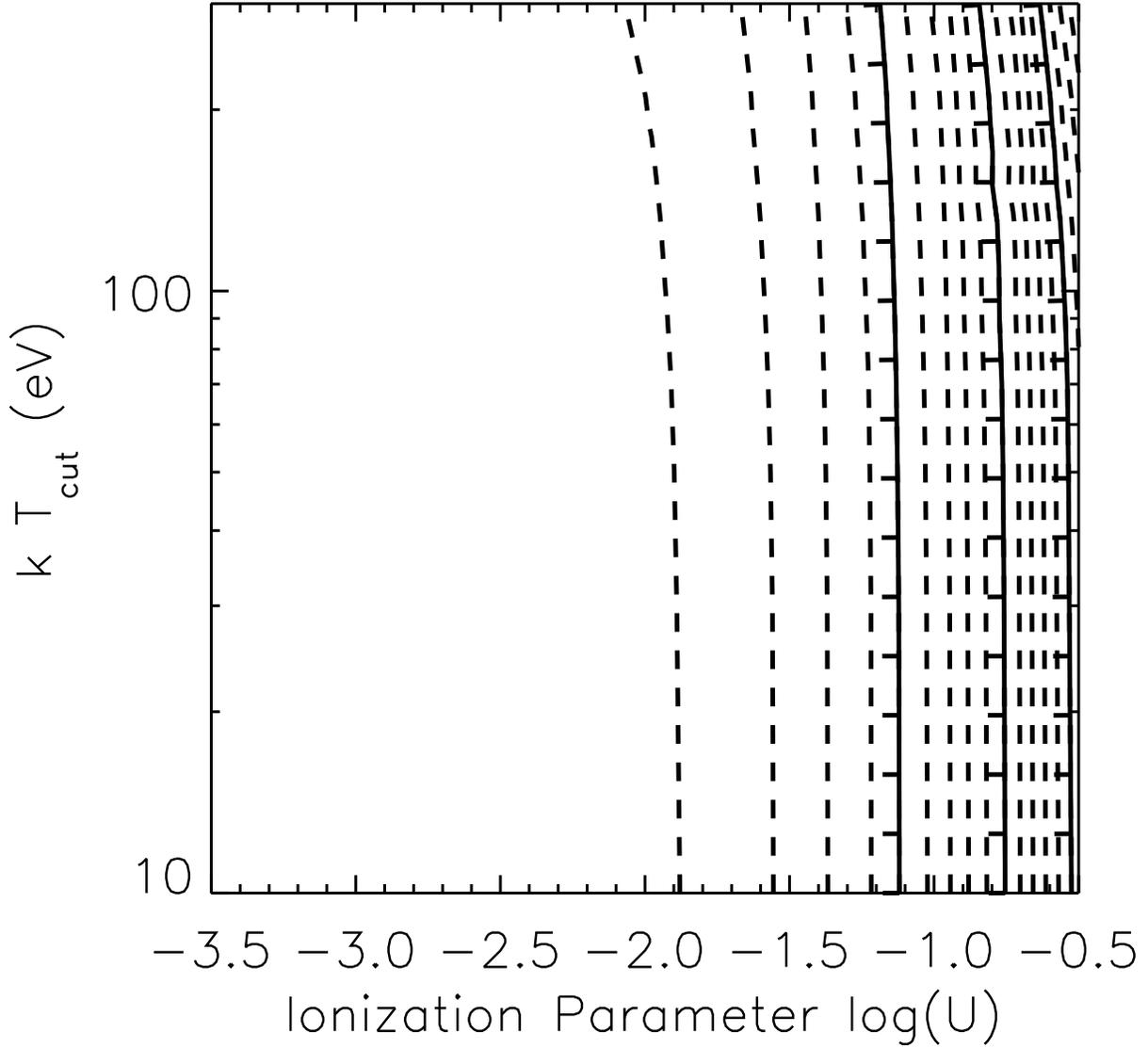}
\caption{\label{fluxLya} The contours of Ly$\alpha$ flux as a function
  of $\log(U)$ and $kT$. This shows that Ly$\alpha$ flux is almost
  independent of spectral energy distribution, as expected.  By
  dividing other line flux values by the Ly$\alpha$, we can partially
  remove the ionization parameter dependence.}

\end{figure}

\clearpage

\begin{figure}
\figurenum{15}
\epsscale{0.4} \plotone{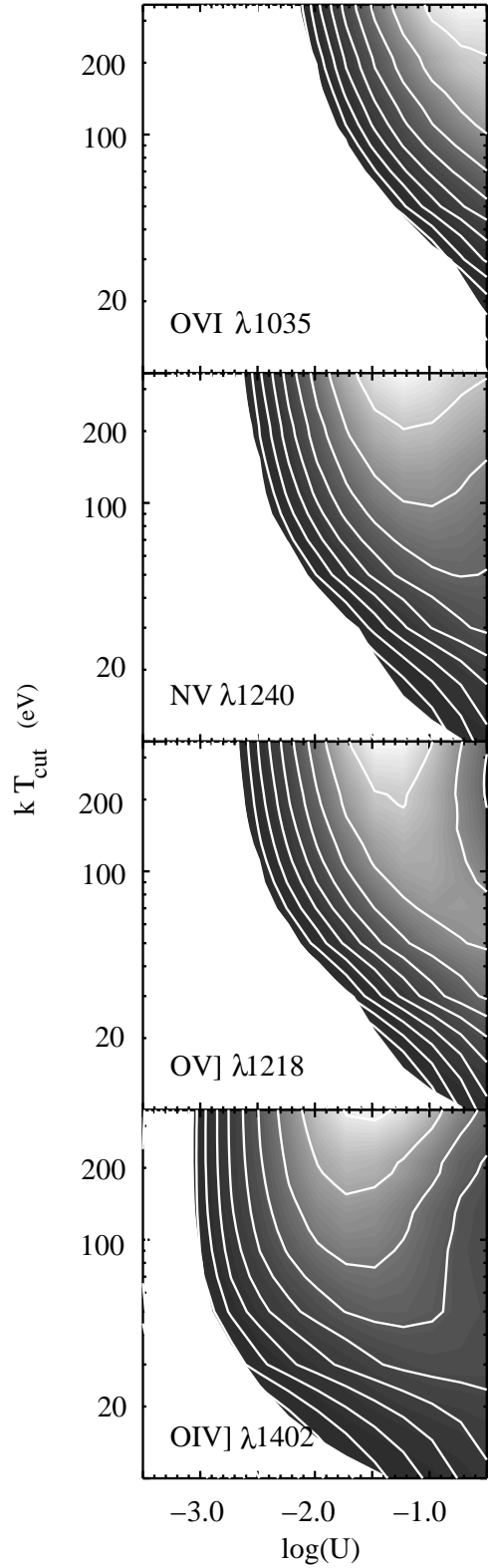}
\caption{\label{HIGHION}Ratios of lines from ions with $\rm I.P. >
  54.4 \, eV$ with Ly$\alpha$. Shading shows the whole range for each
  emission line; contours increase by a factor of 1.25.  The density
  for these simulations is $n_H=10^{10} \rm cm^{-3}$.}
\end{figure}

\clearpage

\begin{figure}
\figurenum{16a}
\epsscale{0.8} \plotone{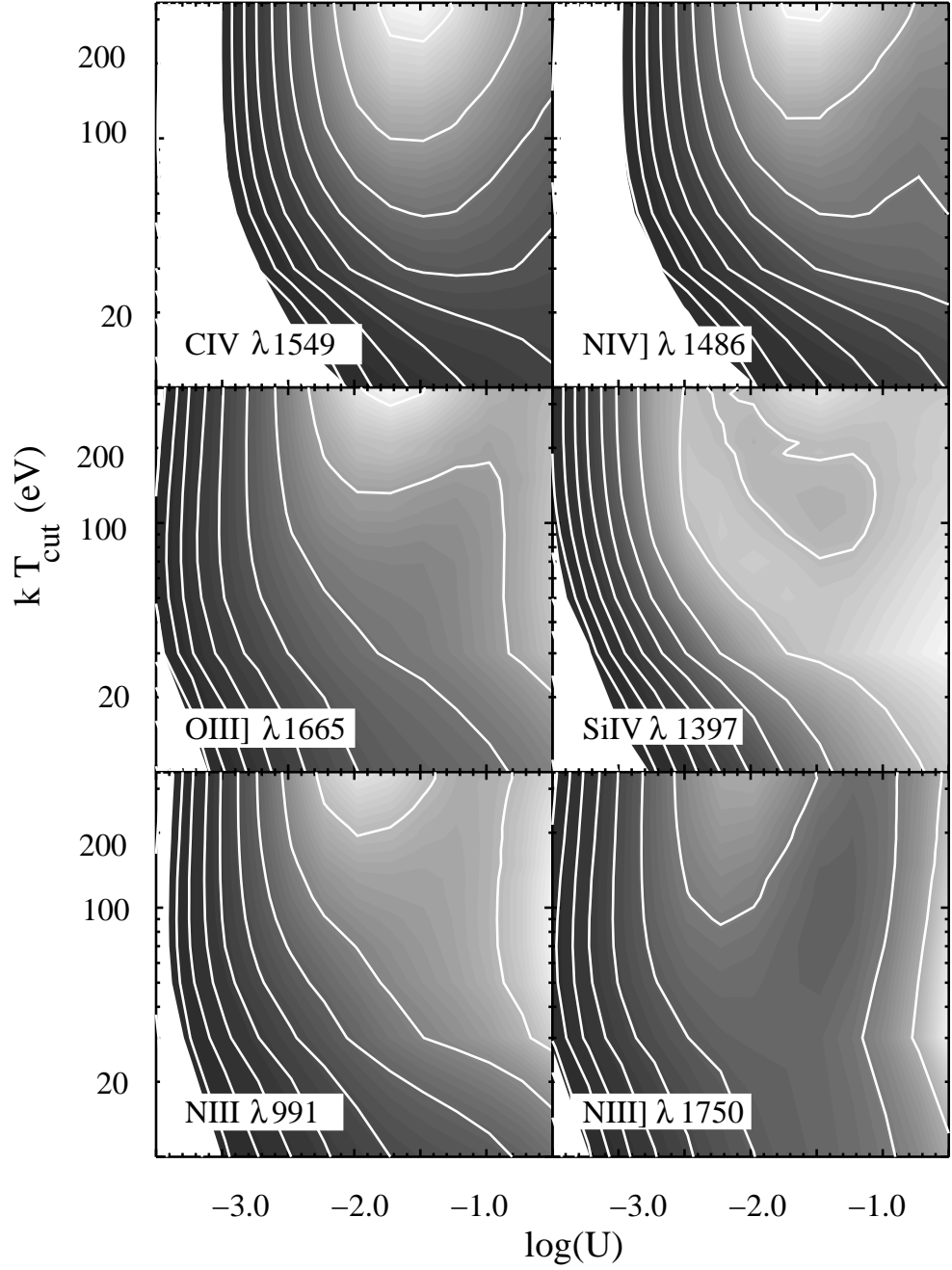}
\caption{\label{MEDION1} Ratios of lines from ions with $\rm 13.6 <
  I.P. < 54.4\,  eV$ with Ly$\alpha$. }
\end{figure}
\clearpage

\begin{figure}
\figurenum{16b}
\epsscale{0.8} \plotone{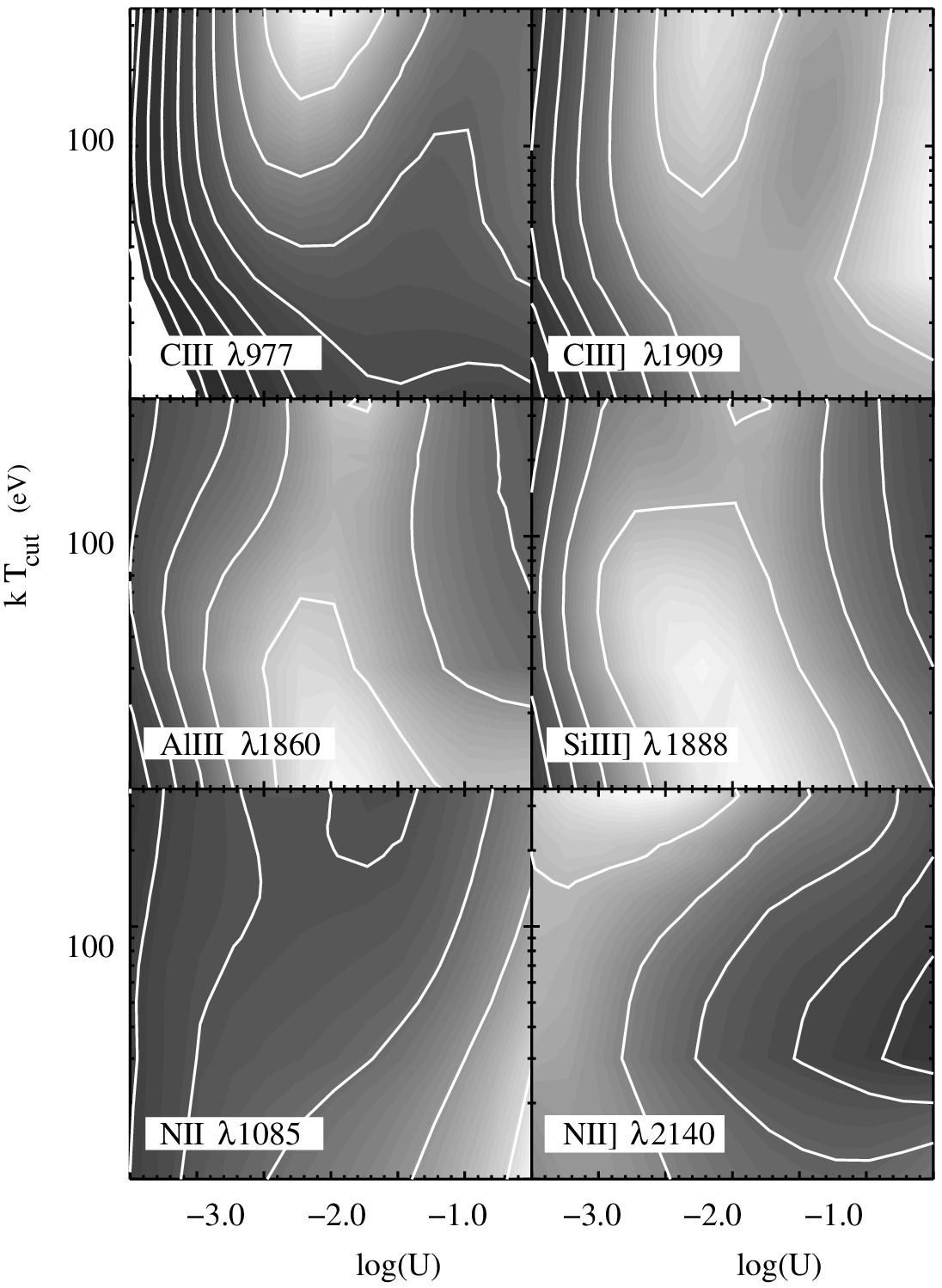}
\caption{\label{MEDION2}Fig.\ 16 continued. }
\end{figure}
\clearpage

\begin{figure}
\figurenum{17}
\epsscale{0.7} \plotone{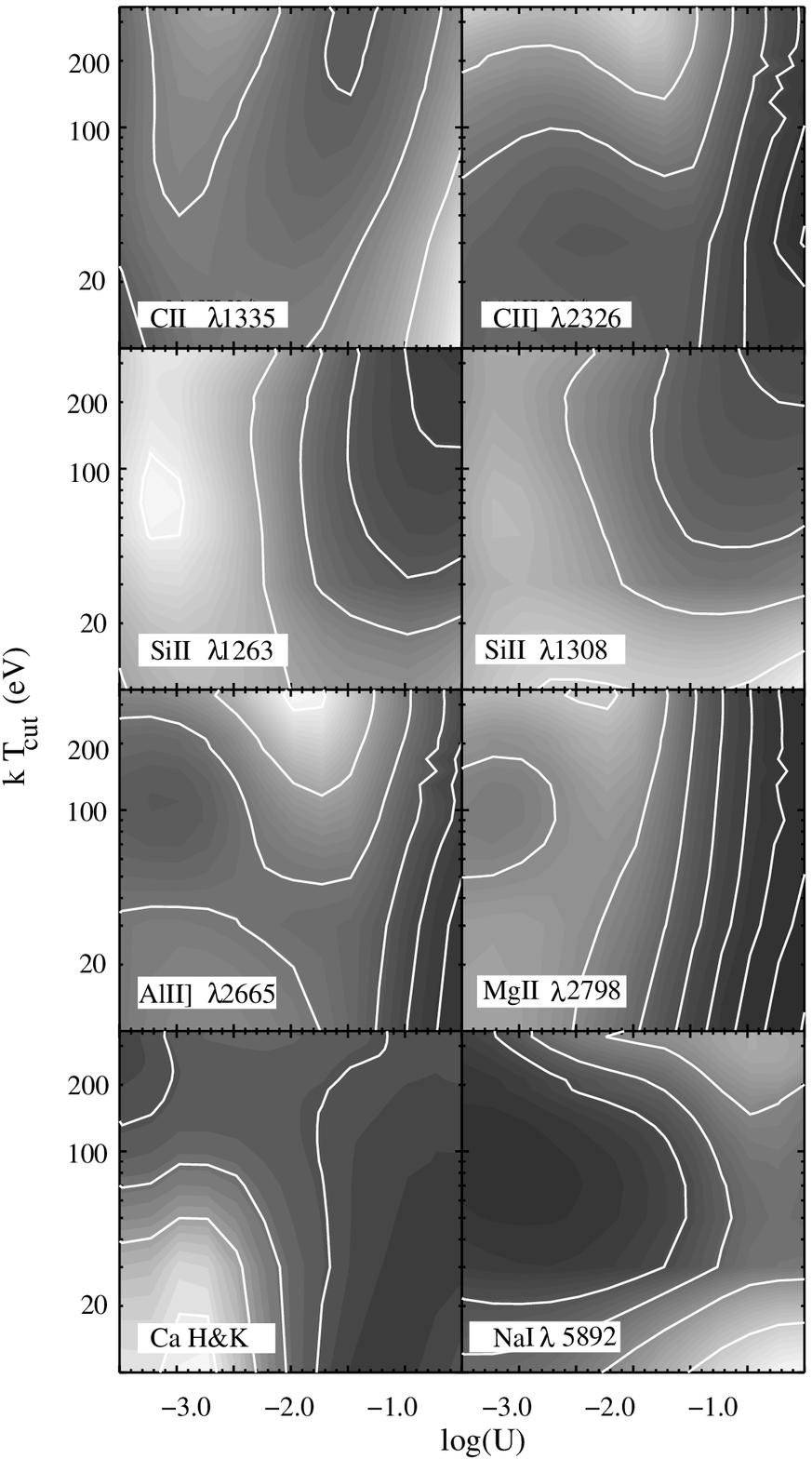}
\caption{\label{LOWION} Ratio of lines from ions with $\rm I.P.<
  13.6\,eV$ with Ly$\alpha$.}
\end{figure}

\clearpage

\begin{figure}
\figurenum{18}
\epsscale{0.7} \plotone{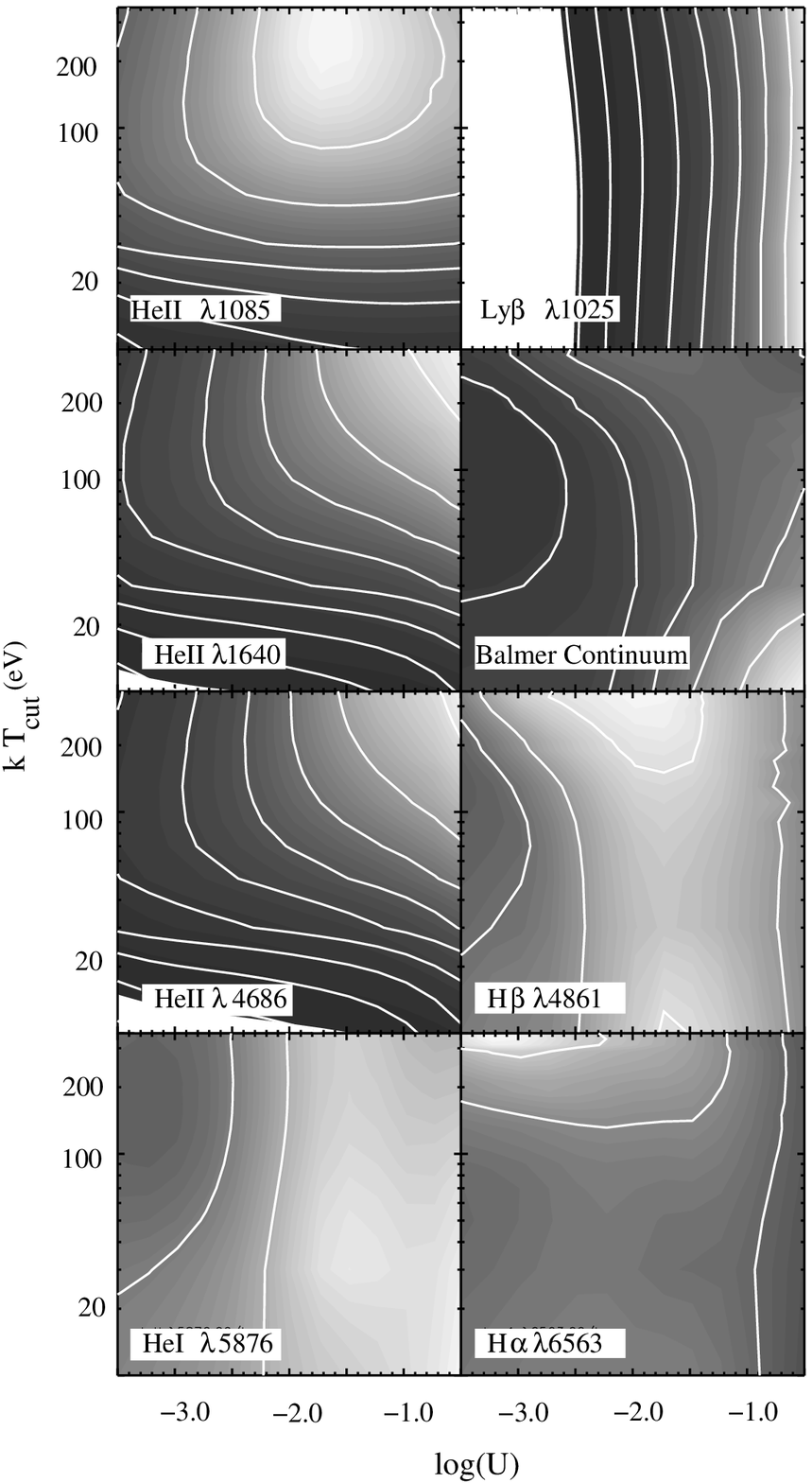}
\caption{\label{RECOMB} Ratio of recombination lines with Ly$\alpha$.}
\end{figure}

\clearpage

\begin{figure}
\figurenum{19}
\epsscale{0.9} \plotone{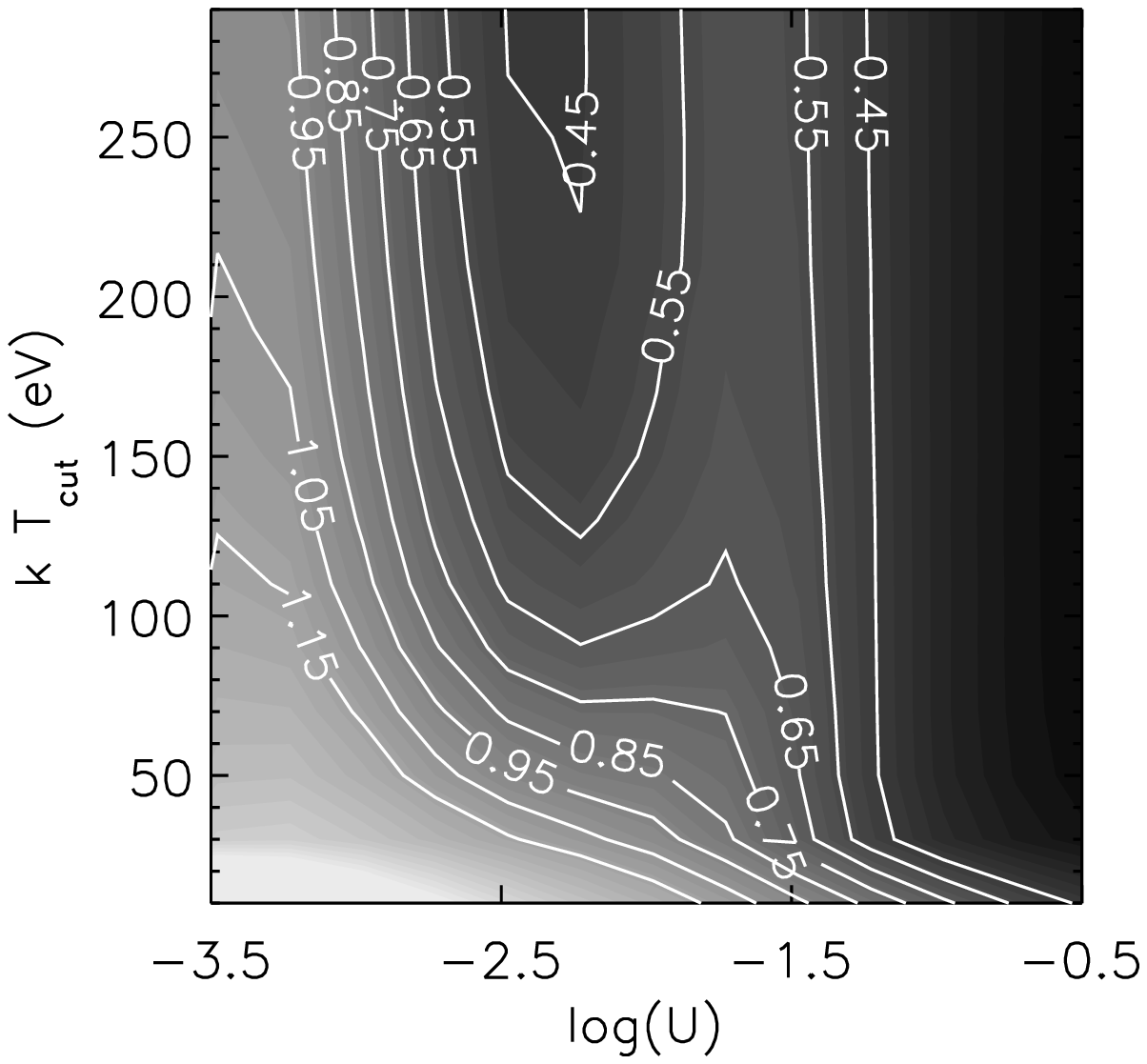}
\caption{\label{DENSITY}Contours for the density indicator
  \ion{Si}{3}]/\ion{C}{3}].  Significant variations in the $kT_{cut}$
  direction indicate that
  \ion{Si}{3}]/\ion{C}{3}] is not independent of the spectral energy
  distribution.  }
\end{figure}

\clearpage

\begin{figure}
\figurenum{20}
\epsscale{1} \plotone{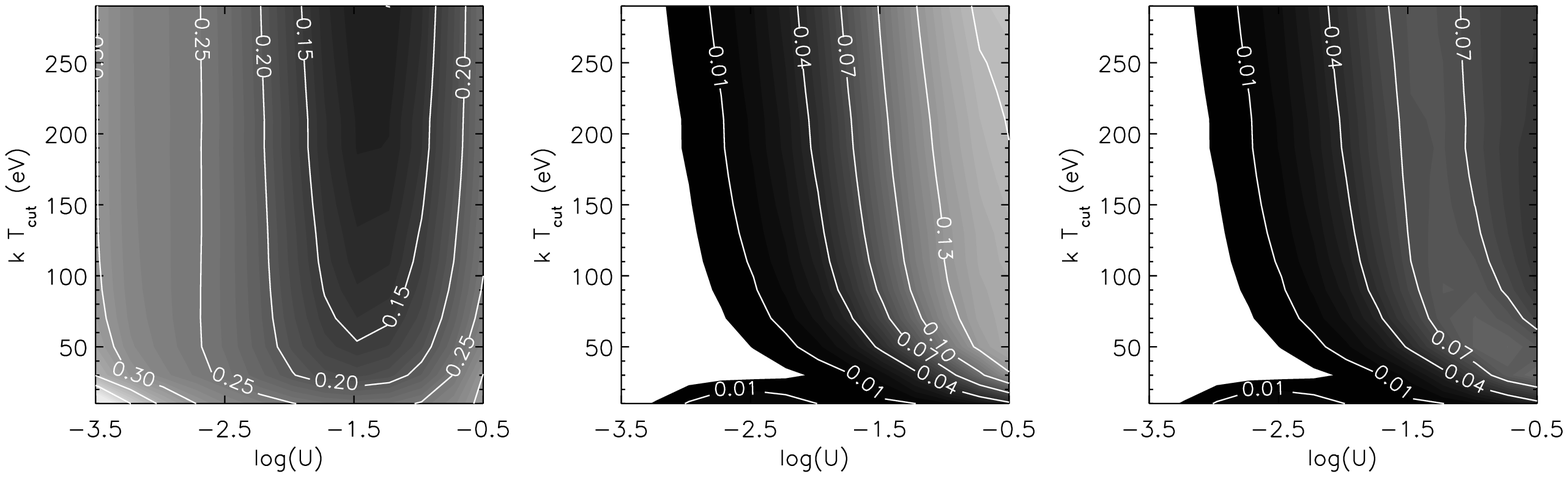}
\caption{\label{METALS1}Contours for the metallicity indicators.
  Left: \ion{N}{3}]/\ion{O}{3}]; middle: \ion{N}{5}/\ion{C}{4}; right:
  \ion{N}{5}/(\ion{O}{6}+\ion{C}{4}).  Variations in the $kT_{cut}$
  direction reveal that the metallicity indicators are not robust for
 the  softest spectral energy distributions.}
\end{figure}

\clearpage

\begin{deluxetable}{lccc}
\tabletypesize{\small}
\tablenum{1}
\tablecaption{Observing Log}
\tablehead{\colhead{Observatory} & \colhead{Date} & \colhead{Bandpass}
  & \colhead{Exposure} \\
& & & \colhead{(seconds)}}
\startdata 
{\it FUSE} & 2000 April 30 20:50:43 -- 2000 May 3 02:32:18  &
905--1187 \AA\/ & 43,640 \\
{\it ASCA} & 2000 May 1 00:14--12:37 & 0.6--10 keV (SIS) & 15,700 \\
 &  & 0.8--10 keV (GIS) & 17,500 \\
{\it EUVE} & 2000 April 1:27:59 -- 2000 May 2 20:57:59  & 0.15 keV &
63,700 \\ 
{\it HST} FOS & 1997 January 31 & G130H (1140--1605) & 4,420 \\
 &  & G190H (1575--2330) & 2,460\\
 &  & G270H (2220--3300) & 670 \\
\enddata
\end{deluxetable}

\clearpage

\begin{deluxetable}{cccccc}
\tablenum{2}
\tablecaption{\label{xray}Spectral parameters in the X-ray band }
\tablewidth{0pt} \tablehead{ \colhead{Correction\tablenotemark{a}}  &
\colhead{Photon index ($\Gamma$)}    & \colhead{PL
Norm.\tablenotemark{b}}   & \colhead{$kT \rm (keV)$ \tablenotemark{c}}   &
\colhead{BB Norm.\tablenotemark{d}}   & \colhead{$\chi^2_\nu$/d.o.f} }
\startdata 
Without & $2.28^{+0.18}_{-0.20}$	&
6.4$^{+1.3}_{-1.2}$	 & 0.152$\pm$0.017	&
$1.3 \pm 0.3$	&1.13/149\\ With	&
$2.35_{-0.16}^{+0.17}$	& $7.1_{-1.1}^{+1.1}$	 &
0.122$\pm$0.014	& $3.4 \pm 0.9$	&1.13/138\\ 
\enddata
\tablenotetext{a}{Correction for SIS data included or not.}  
\tablenotetext{b}{The power-law normalization in units of
  $10^{-4} \rm \, photons\, cm^{-2}\, s^{-1}\, keV^{-1}$ 
at 1~keV in the source rest frame.}  \tablenotetext{c}{The blackbody
temperature in the source rest frame.}  \tablenotetext{d}{Source
luminosity in units of $10^{34}$~ergs~s$^{-1}$ assuming it is located
at 10~kpc. The luminosity distance is 190.2~Mpc using cosmological
parameters of $H_0=70$~km~s$^{-1}$~Mpc$^{-1}$, $\Lambda_0=0.70$, and
$\Omega_M=0.3$.} 
\end{deluxetable}

\clearpage

\begin{deluxetable}{lllrrr}
\tabletypesize{\scriptsize}
\tablenum{3}
\tablecaption{\label{TABLE1}Prominent Emission Line Measurements}
\tablewidth{0pt}

\tablehead{ \colhead{Emission Line} &
  \colhead{measured $\lambda$(\AA)\tablenotemark{a,b,c} }
  & \colhead{Laboratory $\lambda$ (\AA\/)\tablenotemark{a} }
& \colhead{FWHM ($\rm km\,s^{-1}$) \tablenotemark{d} } &
  \colhead{F($10^{-14}\,\rm ergs\, s^{-1} cm^{-2}$)\tablenotemark{c} }
  & \colhead{\ew (\AA\/)}  }  
\startdata 
\ion{C}{3} & $977.4 \pm 0.2$ & 977.0 & $1170 \pm 540$ & $1.6 \pm
  0.3$ &  4.1 \\
Ly$\beta$ & $1028.3 \pm 1.2$ & 1025.7 & $1220 \pm 620$ & $0.20 \pm
  0.12$ &  0.5 \\  
\ion{O}{6} & $1031.7 \pm 0.06$ &  1031.9 & $1300 \pm 140$  &  $11.2
  \pm 1.0$ & 28.8 \\ 
Ly$\alpha$ & $1215.9 \pm 0.02$ & 1215.7 & $790 \pm 20$ &  $22.2 \pm
  0.6$ & 49\\ 
\ion{O}{5}] & $1218.3\pm 0.2$ & 1218.3 & $1020 \pm 130$ & $6.5\pm 0.8$ &  14.4\\ 
\ion{N}{5} & $1239.4 \pm 0.2$ & 1238.8 & $900 \pm 120$ & $2.6 \pm 0.3$
  & 5.8 \\ 
 &  $1243.4 \pm 0.2$ & 1242.8 & $900\pm 120$ & $1.4 \pm 0.2$  & 3.1   \\ 
\ion{C}{2} & $1334.4 \pm 0.3$ & 1334.5 & $300 \pm 90$ & $0.7\pm 0.5 $  & 1.6 \\ 
\ion{O}{4}] & 1399.7 & & $1120 \pm 70$ & $0.09 \pm 0.06$ & 0.23 \\ 
            & 1401.2 & & $1120 \pm 70$ & $0.51 \pm 0.06$ & 1.3 \\ 
            & 1404.8 & & $1120 \pm 70$ & $0.14 \pm 0.06$ & 0.31 \\ 
            & 1407.3 & & $1120 \pm 70$ & $0.09 \pm 0.06$ & 0.20 \\ 
\ion{Si}{4} & $1393.8 \pm 0.3$ & 1393.8 &  $1120 \pm 70$ & $1.33 \pm 0.22$ &3.0  \\
            & $1402.8 \pm 0.3$ & 1402.8 & $1120 \pm 70$ & $1.37 \pm 0.47$  & 3.0 \\ 
\ion{C}{4}  & $1548.0 \pm 0.08$ & 1548.2 &  $900 \pm 30$ & $7.9 \pm
  0.4$ & 14.4 \\ 
            & $1550.5 \pm 0.08$ & 1550.8 & $900 \pm 30$  & $8.4 \pm
  0.5$ &  15.2 \\ 
\ion{He}{2} (blueshifted) & $1637 \pm 0.1$\tablenotemark{e} & 1640.4 &
  $700 \pm 140$ & $1.0 \pm 0.2$    & 1.8 \\  
\ion{He}{2}  & $1640.7 \pm 0.1$ & 1640.4 & $570 \pm 60$ & $2.7 \pm 0.3$ & 5.0 \\ 
\ion{O}{3}] & $1661.0 \pm 1.4$  & 1660.8 & $810 \pm 40$ & $0.45 \pm
  0.21$ &  0.8\\ 
            & $1666.3 \pm 0.7$ & 1666.2 & $810 \pm 40$  & $1.0 \pm 0.18$ &  1.8\\ 
\ion{N}{3}] & $1751.0 \pm 0.5$  & 1750.4 & $810 \pm 40$   & $0.62 \pm
  0.13$ & 1.1 \\ 
\ion{Si}{3}] & $1892.3 \pm 0.3$ & 1892.0 & $760 \pm 130$ & $1.18 \pm
  0.15$ & 2.1 \\  
\ion{C}{3}] & $1909.1 \pm 0.08 $ & 1908.7 & $810 \pm 40$ & $4.45 \pm 0.16$ & 8.1 \\ 
\ion{Mg}{2} &  $2796.4 \pm 0.3$ & 2796.4 & $740 \pm 70$ & $1.0 \pm 0.09$ & 5.0 \\ 
            & $2803.6 \pm 0.3$ & 2803.5 &  $740 \pm 70$  & $1.4 \pm 0.10$ &  7.0\\ 
\enddata
\tablenotetext{a}{Vacuum wavelengths for $\lambda < 3000$\AA\/; Air
            wavelengths for $\lambda > 3000$\AA\/.}
\tablenotetext{b}{No entry in this column means that the wavelengths
  were fixed at the laboratory wavelength for the purpose of
  deblending weak lines.}
\tablenotetext{c}{Repeated uncertainties in these column mean that the
  flux ratios were constrained in the fitting.}
\tablenotetext{d}{Repeated values in this column mean that the
  widths were constrained to be equal in the  fitting.}
\tablenotetext{e}{A blueshifted component was observed on 
            \ion{He}{2}; this is the central wavelength of that component.}
\end{deluxetable}

\clearpage

\begin{deluxetable}{lllll}
\tablenum{4}
\tablecaption{Emission Line Comparison}
\tablewidth{0pt}
\tablehead{
\colhead{Emission line$^a$} & \colhead{RE 1034$+$39} &
\multicolumn{3}{c}{Relative Flux}\\
 & \colhead{Relative Flux} & \multicolumn{3}{c}{\hrulefill} \\
& & \colhead{Francis} & \colhead{Zheng$^a$} & \colhead{Brotherton}}

\startdata
Ly$\beta$+\ion{O}{6} & 102\tablenotemark{c} & 9.3 & 16.0 & 11.6 \\
Ly$\alpha$+\ion{N}{5} feature & 118 & 100 & & 100 \\
Ly$\alpha$  & 100 & & 100 & \\
\ion{N}{5}  & 18.0 & & 14.0 &  \\
1400 \AA\ feature & 15.9 & 19 & 8.1 & 6.8 \\
\ion{He}{2} & 16.8\tablenotemark{b}  & 18\tablenotemark{d} & 3.4 &
5.2\tablenotemark{d} \\
\ion{C}{4} & 73 & 63 & 52 & 27 \\
1900 \AA\ feature & 25.4 & 29 & & 10 \\
\ion{Al}{3} & 0.0 & & 2.9 &  \\
\ion{Si}{3}] & 5.3 & & 2.8 &  \\
\ion{C}{3}] & 20.0 & & 12 &  \\
\ion{Mg}{2} & 10.8 & 34 & 33 & 13 \\
\enddata

\tablenotetext{a}{Radio-quiet quasars.}
\tablenotetext{b}{Sum of blueshifted and rest frame components..}
\tablenotetext{c}{Assuming that the 1037\AA\/ and 1031\AA\/ lines have
  the same flux (optically thick). The value would be 77 if the lines
  have 2:1 ratio (optically thin).}
\tablenotetext{d}{\ion{He}{2}~$\lambda 1640$+\ion{O}{3}]~$\lambda 1666$.}
\end{deluxetable}

\clearpage

\begin{deluxetable}{lcccc}
\tablenum{5}
\tabletypesize{\small}
\tablecaption{\label{LOC}LOC model results} \tablewidth{0pt}

\tablehead{
\colhead{Emission Line} & 
\colhead{Baldwin \tablenotemark{a,b}} & 
\colhead{Baldwin Re-created\tablenotemark{a,c}} & 
\colhead{\re SED \tablenotemark{a}}  &
\colhead{RE~1034$+$39 (measured)\tablenotemark{a}}}

\startdata
\ion{O}{6}~$\lambda 1032$ + Ly$\beta$  & 0.16  & 0.19  & 0.52 & 0.51 \\ 
Ly$\alpha$    & 1.0   & 1.0 & 1.0  & 1.0 \\ 
\ion{N}{5}   & 0.04  & 0.067 & 0.18  & 0.18\\ 
\ion{O}{4}]+\ion{Si}{4}   & 0.06 & 0.05  & 0.17  &  0.16 \\ 
\ion{C}{4}    & 0.57  & 0.54  & 1.11& 0.73 \\ 
\ion{O}{3}] + \ion{He}{2}~$\lambda 1640$  & 0.14 & 0.16  & 0.25 & 0.11\tablenotemark{d} \\ 
\ion{Mg}{2}  & 0.34  & 0.29  & 0.47 & 0.11\\ 
\enddata

\tablenotetext{a}{Flux ratios with respect to Ly$\alpha$.}
\tablenotetext{b}{The results taken from Balwin et al.\ 1995.}
\tablenotetext{c}{Our re-creation of the Balwin et al.\ (1995) results.}
\tablenotetext{d}{Only narrow, restframe component of \ion{He}{2} is
  included.} 
\end{deluxetable}

\end{document}